\documentclass[superscriptaddress,twocolumn,prx,preprintnumbers,amsmath,amssymb]{revtex4-2}
\usepackage{braket}
\usepackage{bm}
\usepackage{graphicx,afterpage}
\usepackage{xcolor}
\usepackage{ulem}
\usepackage[colorlinks=true,plainpages=false,linkcolor=blue,urlcolor=blue,citecolor=blue,pdfpagemode=UseNone,pdfstartview=FitBH]
{hyperref}
\usepackage{amsmath} 
\newcommand{\angstrom}{\text{\normalfont\AA}}
\newcommand{\ybco}{YBa$_{2}$Cu$_{3}$O$_{\rm{y}}$}
\usepackage[misc]{ifsym} 
 
\newcommand{\mycomment}[1]{}

\def\bPhi{{\bm \Phi}}
\def\bnabla{\bm{\nabla}}
\def\rhos{\rho_s}

\def\br{{\bf r}}

\def\bq{{\bf q}}
\def\bQ{{\bf Q}}

\def\bQ{{\bf Q}}

\def\tU{\tilde{U}}
\def\tJ{\tilde{J}}

\begin{document}

\title{Using strain to uncover the interplay between two- and three-dimensional charge density waves in high-temperature superconducting \ybco}

\author{I. Vinograd}
\altaffiliation{These authors contributed equally to this work}
\affiliation{Institute for Quantum Materials and Technologies, Karlsruhe Institute of Technology, Kaiserstr. 12, D-76131 Karlsruhe, Germany}
\affiliation{4th Physical Institute – Solids and Nanostructures, University of Göttingen, D-37077 Göttingen, Germany}
\author{S. M. Souliou}
\altaffiliation{These authors contributed equally to this work}
\affiliation{Institute for Quantum Materials and Technologies, Karlsruhe Institute of Technology, Kaiserstr. 12, D-76131 Karlsruhe, Germany}
\author{A.-A. Haghighirad}
\affiliation{Institute for Quantum Materials and Technologies, Karlsruhe Institute of Technology, Kaiserstr. 12, D-76131 Karlsruhe, Germany}
\author{T. Lacmann}
\affiliation{Institute for Quantum Materials and Technologies, Karlsruhe Institute of Technology, Kaiserstr. 12, D-76131 Karlsruhe, Germany}
\author{Y. Caplan}
\affiliation{Racah Institute of Physics, The Hebrew University, Jerusalem 91904, Israel}
\author{M. Frachet}
\affiliation{Institute for Quantum Materials and Technologies, Karlsruhe Institute of Technology, Kaiserstr. 12, D-76131 Karlsruhe, Germany}
\author{M. Merz}
\affiliation{Institute for Quantum Materials and Technologies, Karlsruhe Institute of Technology, Kaiserstr. 12, D-76131 Karlsruhe, Germany}
\affiliation{Karlsruhe Nano Micro Facility (KNMFi), Karlsruhe Institute of Technology, Kaiserstr. 12, D-76131 Karlsruhe, Germany}
\author{G. Garbarino}
\affiliation{ESRF, The European Synchrotron, 71, avenue des Martyrs, CS 40220 F-38043 Grenoble Cedex 9}
\author{Y. Liu}
\affiliation{Max Planck Institute for Solid State Research, Heisenbergstraße 1, D-70569 Stuttgart, Germany}
\author{S. Nakata}
\affiliation{Max Planck Institute for Solid State Research, Heisenbergstraße 1, D-70569 Stuttgart, Germany}
\author{K. Ishida}
\altaffiliation{Current address: Institute for Materials Research, Tohoku University, Sendai 980-8577, Japan}
\affiliation{Max Planck Institute for Chemical Physics of Solids, N\"{o}thnitzer Str. 40, D-01187 Dresden, Germany}
\author{H. M. L. Noad}
\affiliation{Max Planck Institute for Chemical Physics of Solids, N\"{o}thnitzer Str. 40, D-01187 Dresden, Germany}
\author{M. Minola}
\affiliation{Max Planck Institute for Solid State Research, Heisenbergstraße 1, D-70569 Stuttgart, Germany}
\author{B. Keimer}
\affiliation{Max Planck Institute for Solid State Research, Heisenbergstraße 1, D-70569 Stuttgart, Germany}
\author{D. Orgad}
\affiliation{Racah Institute of Physics, The Hebrew University, Jerusalem 91904, Israel}
\author{C. W. Hicks}
\affiliation{Max Planck Institute for Chemical Physics of Solids, N\"{o}thnitzer Str. 40, D-01187 Dresden, Germany}
\affiliation{School of Physics and Astronomy, University of Birmingham, Birmingham, B15 2TT, UK}
\author{M. Le Tacon}
\thanks{Correspondance and requests for materials to \href{mailto:matthieu.letacon@kit.edu}{matthieu.letacon@kit.edu}}
\affiliation{Institute for Quantum Materials and Technologies, Karlsruhe Institute of Technology, Kaiserstr. 12, D-76131 Karlsruhe, Germany}

\begin{abstract}
Uniaxial pressure provides an efficient approach to control the competition between charge density waves (CDWs) and superconductivity in underdoped \ybco. It can enhance the correlation volume of ubiquitous short-range 2D CDW correlations, and induces a long-range 3D CDW otherwise only accessible at large magnetic fields. Here, we use x-ray diffraction to study the strain and doping evolution of these CDWs. No signatures of discommensurations nor pair density waves are observed in the investigated strain-temperature parameter space, but direct evidence for a form of competition between 2D and 3D CDWs is uncovered. We show that the interplay between the 3D CDW, the 2D CDWs and superconductivity is qualitatively well described by including strain effects in simulations of a nonlinear sigma model of competing superconducting and CDW orders. From a broader perspective, our results underscore the potential of strain tuning as a powerful tool for probing and manipulating competing orders in quantum materials.
\end{abstract}

\maketitle

\section{Introduction}
A tendency towards charge ordering in the underdoped high-temperature superconducting cuprates was predicted soon after their discovery~\cite{Zaanen_PRB1989,Poilblanc1989, Emery_PRL1990}. However, it took decades of effort with a variety of experimental methods to demonstrate this phenomenon and its ubiquity across the cuprate superconductors~\cite{Tranquada_Nature1995, Howald_PRB2003, Hoffman_Science02_1, Wise_NaturePhysics2008, Wu_Nature2011, Ghiringhelli2012, Chang2012, Tabis_NatCom2014, daSilvaNeto_Science2014}. This effort has raised many new questions. One observation that lacks a clear explanation is the fact that the competition between charge order and superconductivity is striking in some cuprate families~\cite{Ghiringhelli2012, Chang2012, Croft_PRB2014}, but is less pronounced in others~\cite{Tabis_NatCom2014, daSilvaNeto_Science2014,Tabis_PRB2017}. Another is the very substantial variation in correlation length, which varies over two orders of magnitude amongst the cuprates families --- from a few unit cells to several tens of nm --- without any
obvious relation to the superconducting critical temperature $T_\text{c}$. To address these issues, it has proved useful to focus on materials with low disorder scattering, and to probe them with external tuning parameters, such as magnetic field~\cite{Chang2012} or pressure~\cite{Souliou2018, Vinograd2019,Souliou2020}, that do not introduce disorder. While hydrostatic pressure has been widely used to investigate unconventional superconductors and correlated electron systems, recent studies have highlighted several benefits of uniaxial stress. In the cuprates, uniaxial stress has been used to probe charge stripes in La-based compounds~\cite{Boyle2021, Wang2022, gupta2023tuning}, and to polarize them~\cite{Simutis2022,Jakovac2023}. In YBa$_2$Cu$_3$O$_{6.67}$, substantial suppression of $T_\text{c}$ has been achieved with uniaxial stress~\cite{Barber2022}, and in contrast 
to the introduction of pair-breaking impurities~\cite{Blanco_PRL2013} or magnetic fields~\cite{Chang2012} (which introduce vortices), the suppression is homogeneous. Suppressing $T_\text{c}$ allows
alternative electronic orders to develop, and homogeneity allows the interplay between these orders and the superconductivity to be studied with precision.

\begin{figure}
\hspace*{-0.0cm}    
  \includegraphics[width=8.5cm]{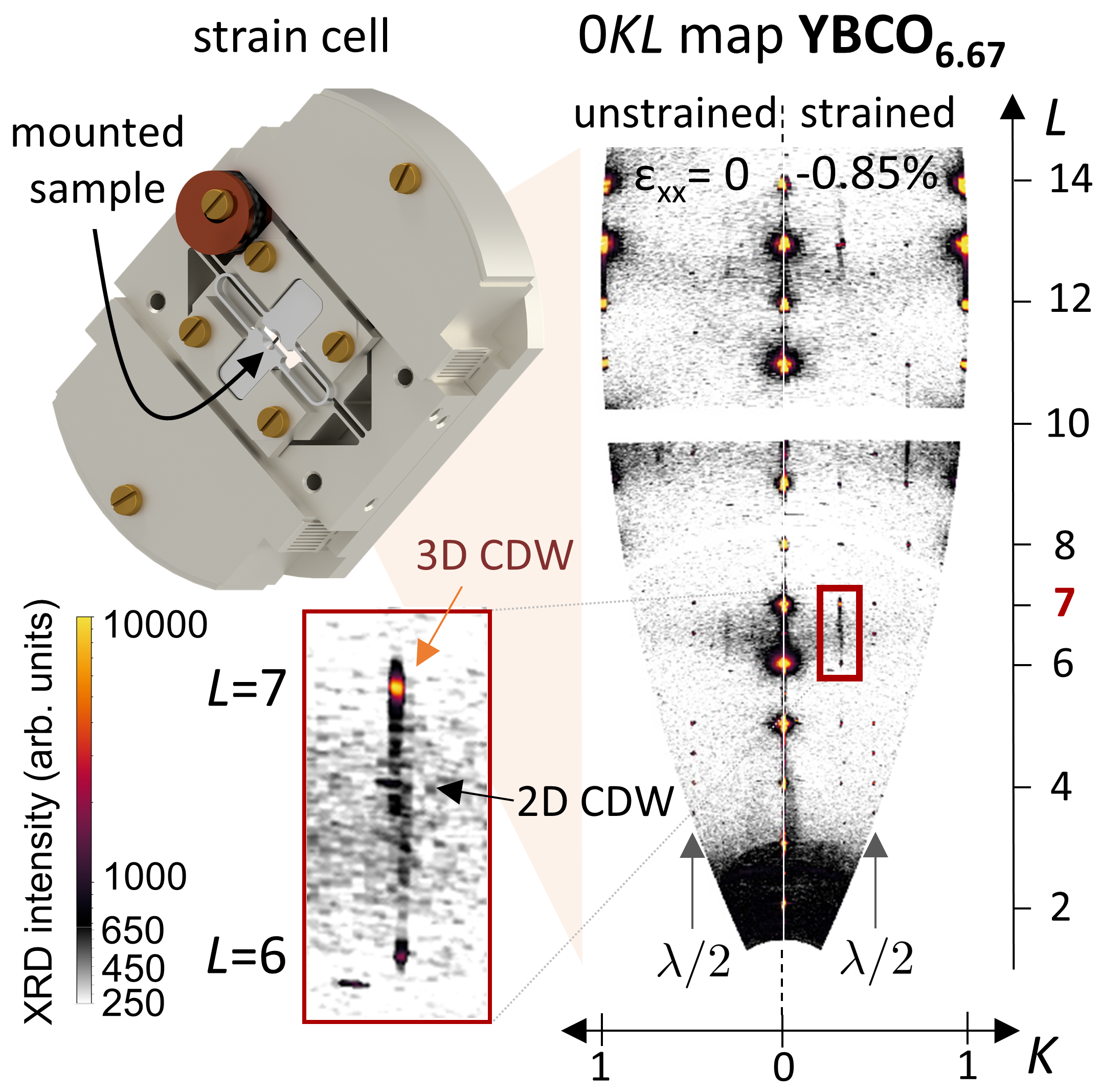}
  \caption{\label{strain_cell_0KL_6p67} \textbf{Experimental setup and representative results of x-ray scattering under uniaxial stress.}  Measurements were performed in transmission geometry using a Razorbill CS200T strain cell. The YBCO$_{6.67}$ sample was mounted onto an exchangeable titanium cross. Sections of large ($0KL$) reciprocal space maps at zero (left) and highest $a$ axis compression (right) show the appearance of sharp 3D CDW peaks at integer $L$. Faint 2D CDW intensity in between is present at all strains at half-integer $L$ on top of a diffuse phonon background. Grey arrows mark sharp Bragg reflections that appear at $K=0.5$ due to higher harmonics with $\lambda/2$. The inset shows an enlarged part of the $(0KL)$ map. To make the 2D CDW more apparent, the colour scale is linear for small and logarithmic for large intensities.}
\end{figure}

YBa$_2$Cu$_3$O$_{6.67}$ is centered within a doping range where an incommensurate, short-range, 2D-correlated charge order (hereafter referred to as the 2D CDW) is
strongest~\cite{Blanco2014}. Uniaxial compression along the $a$ axis (the shortest axis of the orthorhombic structure) 
has been shown to suppress $T_\text{c}$ by up to
30\%~\cite{Barber2022}, and enhance the amplitude and correlation lengths of the 2D CDW~\cite{Kim2018, Kim2021}. Beyond a threshold stress value, a long-range-ordered, 3D-correlated CDW (hereafter referred to as the 3D CDW) emerges~\cite{Kim2018, Kim2021}. 3D CDW order can also be induced by a magnetic field, though to date, with lower amplitude and shorter correlation lengths than the one induced by uniaxial stress~\cite{Wu_Nature2011,Wu2013,Gerber2015,Chang2016,Jang2016,Choi2020}. The 2D CDW is biaxial, with components that propagate along both the $a$ and $b$ axes, while the 3D CDW is uniaxial, propagating along the $b$ axis only. 
On the other hand, it the in-plane wavelengths of the 2D and 3D CDWs are identical~\cite{Kim2018,Gerber2015} and the analysis of the in-plane correlation lengths revealed that the 2D CDW consists of quasi-independent unidirectional orders~\cite{Comin_Science2015}. This has further been confirmed by the observation of substantially different dependencies of amplitude and correlation lengths of its $a$ and $b$ components on uniaxial stress~\cite{Kim2021}. Overall, this suggests that the 2D and 3D CDWs might be intimately related.

A key question remains to be clarified: does the 3D CDW emerge from ``patches'' of 2D CDWs that lock together, or is it a separate order, formed by different charge carriers, whose periodicity locks to that of the background 2D CDWs? Additionally, so far there is little information on where the 3D CDW exists in strain-temperature space, which is crucial information for understanding its interaction with the superconductivity. Here, we study underdoped YBa$_2$Cu$_3$O$_{\rm{y}}$ under uniaxial stress with synchrotron hard x-ray diffraction. We show that the 2D CDW amplitude stops growing when the 3D CDW onsets, and map the boundaries of the 3D CDW phase in a strain-temperature phase diagram. 
We further show that the strain-induced formation of the 3D CDW and its interplay with the 2D CDW and superconductivity are qualitatively well described by including the effects of strain in an extension of a nonlinear sigma model~\cite{Hayward_Science2014} previously used to investigate the magnetic-field-induced formation of the long-range 3D order in YBCO~\cite{Caplan2015,Caplan2017}. In this picture, the long-range 3D CDW emerges from the regions between disorder-induced 2D CDW domains (\textit{i.e.}, the regions where superconductivity forms in the absence of strain). The growth of disordered 2D CDW domains is then prevented against increased surface tension due to phase-mismatch at the boundary with phase-ordered 3D CDW.
Finally, we have also looked for signs of discommensurations, which would indicate a tendency of the CDW to lock to the lattice, and for pair density wave correlations on top of the 3D CDW, but have found no evidence for either.

\begin{figure}
\hspace*{-0.2cm}    
  \includegraphics[width=8.5cm]{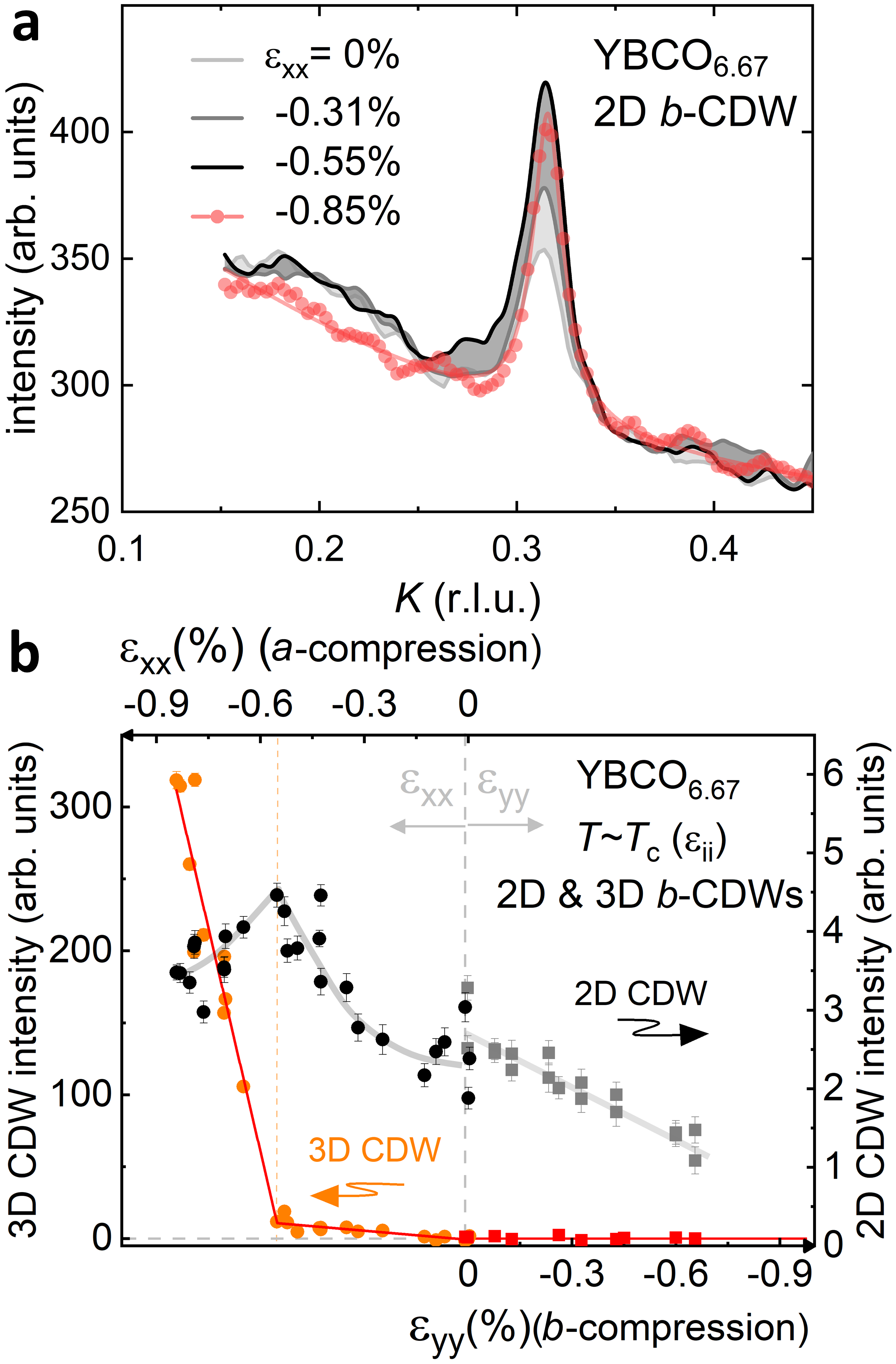}
  \caption{\label{2D_and_3D_plots} \textbf{Strain dependence of 2D and 3D CDW intensity for YBCO$_{6.67}$.} a) 2D $b$-CDW intensity from $K$-cuts at half-integer $L$ values for $a$ axis compression. $a$ axis compression amplifies the $b$-CDW only up to the onset of 3D ordering.
  b) $a$ axis and $b$ axis compression dependence of the integrated 2D CDW and 3D CDW intensities from $K$-cuts at half-integer $L$=6.5 and $L$-cuts through $q_{\rm{CDW}}$, respectively. The vertical scales of the integrated intensities (line cuts) at $T\sim T_c(\varepsilon_{xx})$ correspond to the same total integrated intensity, $I_{tot}$, based on Supplementary Table~II in Supplementary Note 2. The orange dashed line marks the onset of long-range 3D CDWs at $\varepsilon_{xx}$=-0.55\%.}
\end{figure} 

\begin{figure}[t!!!]
\hspace*{-0.2cm}    
  \includegraphics[width=8cm]{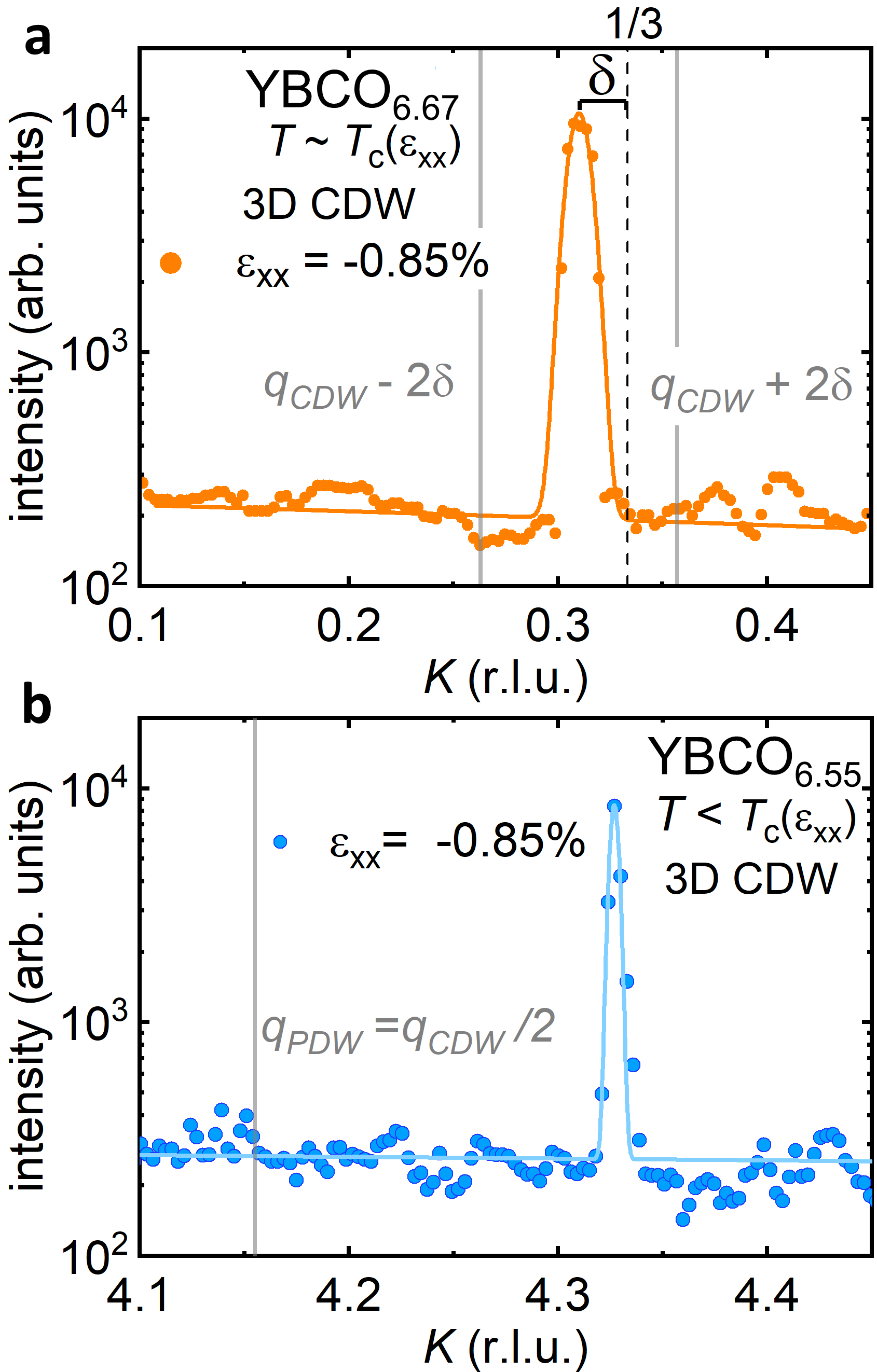}
  \caption{\label{discommensuration} \textbf{$K$-cuts of the strain-induced 3D CDW.} a) YBCO$_{6.67}$ $K$-cuts through the 3D CDW peak at integer $L$, maximal compression and $T=60.5$~K~$\sim T_{\rm{c}}$ show an intense peak at incommensurate $q_{\rm{CDW}}=$~0.310(1) r.l.u., with $\delta$ marking the incommensurability with respect to the closest commensurate value, 1/3. Vertical grey lines mark where discommensurations would lead to satellite peaks at $q_{\rm{CDW}}\pm 2\delta$. b) YBCO$_{6.55}$ $K$-cut through the 3D CDW peak at integer $L$, maximal compression and $T=45$~K~$< T_{\rm{c}}$ in the superconducting state, where the vertical grey line marks where pair density waves are expected to produce a peak at $q_{\rm{PDW}}=q_{\rm{CDW}}/2=$~0.327(3) r.l.u./2. }
\end{figure}

\begin{figure*}[t!!!]
\hspace*{-0.0cm}    
  \includegraphics[width=0.99\textwidth]{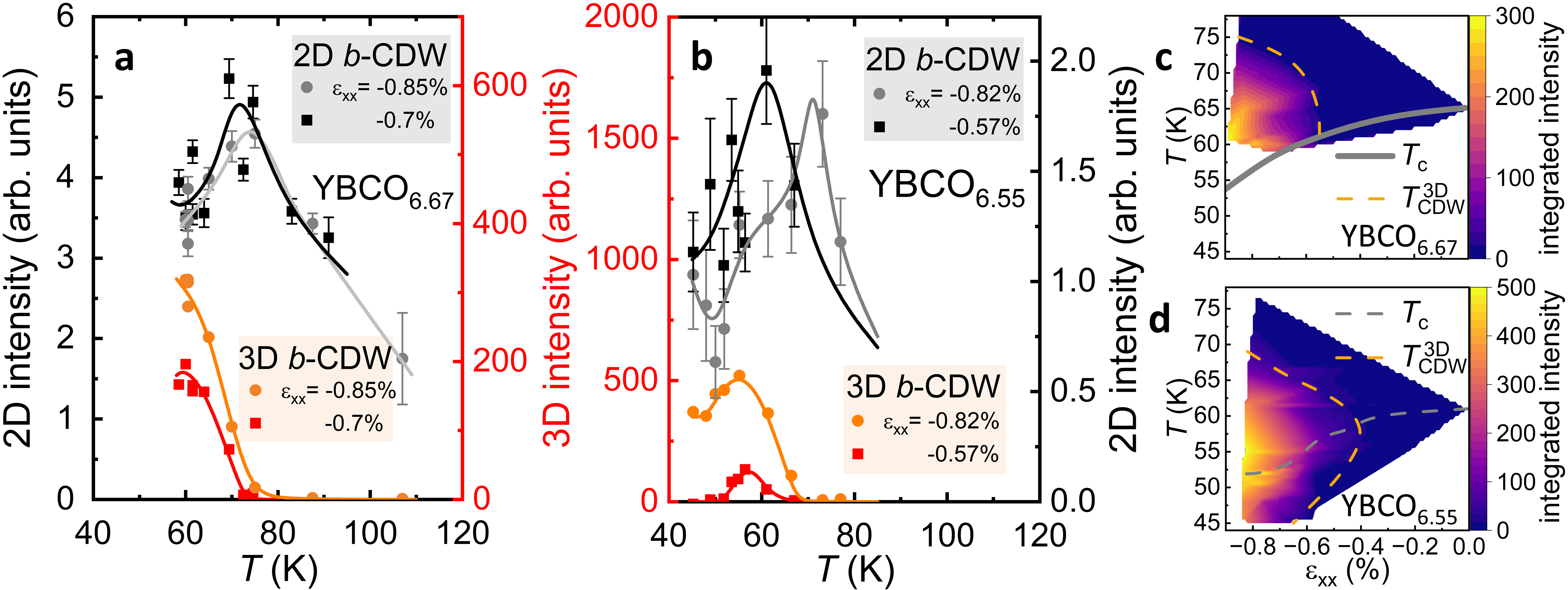}
  \caption{\label{temperature dependence} \textbf{Temperature dependence of 2D and 3D CDW intensity for YBCO$_{6.67}$ and YBCO$_{6.55}$.} a) Temperature dependence of 3D and 2D CDW intensities for YBCO$_{6.67}$ at $a$ axis compressions $\varepsilon_{xx}$=-0.7~\% and -0.85~\%. b) 3D and 2D CDW intensities for YBCO$_{6.55}$ at compressions of $\varepsilon_{xx}$=-0.57~\% and -0.82~\%. 2D CDW (3D CDW) intensities are determined from $K$-cuts ($L$-cuts). Lines are guides to the eye. For a comparison of 2D and 3D intensities, see Table~\ref{tab:correlation length}. c) Strain-temperature phase diagram of the 3D CDW intensity for YBCO$_{6.67}$ with the $T_c$ curve reproduced from ref.~\cite{Barber2022}. Dashed orange lines mark the boundary of the 3D CDW phase. In the corresponding diagram for YBCO$_{6.55}$ in panel d) $T_c(\varepsilon_{xx})$ is estimated from the linear extrapolation down to the onset of the 3D phase~\cite{Kraut1993}. Within the 3D CDW phase $T_c(\varepsilon_{xx})$ is based on the peak in the 3D CDW intensity.}
\end{figure*}

\section{Results}

{\bf Experimental results.} Hard x-ray diffraction was performed as a function of temperature on uniaxially pressurized \ybco single crystals, as described in the Methods section. We first discuss the effect of uniaxial compression along the $a$ and $b$ directions on the 2D CDW, for samples with $\rm{y} = 6.67$ ($p = 0.125$).  As shown in the intensity maps in Fig.~\ref{strain_cell_0KL_6p67}, the $b$ component of the 2D CDW has an extremely broad profile along the $[0,0,L]$ direction, centered at half-integer $L$, but is relatively sharp along the $[0,K,0]$ direction already in the absence of strain. The incommensurate wavevector of the modulations are in perfect agreement with previous reports~\cite{Ghiringhelli2012, Chang2012, Blanco2014, Blanco_PRL2013}.
To evaluate intensity along the $K$ direction, we integrate the intensity over the range $0.4<L<0.6$, with results at a few strains shown in Fig.~\ref{2D_and_3D_plots}a. For these measurements, the temperature is close to $T_\text{c}(\varepsilon_{xx})$. For strains not exceeding the 3D CDW onset strain, the 2D CDW is most intense at $T \approx T_\text{c}(\varepsilon_{xx})$. It can be seen in Fig.~\ref{2D_and_3D_plots}a that the intensity of the $b$ component of the 2D CDW approximately doubles between $\varepsilon_{xx} = 0$ and $-0.55$\% (but the CDW ordering wave vector is only weakly affected by strain). 
The integrated intensity at a denser set of strains is shown in Fig.~\ref{2D_and_3D_plots}b, in which this doubling is again visible, along with a reduction in intensity for $\varepsilon_{xx} < -0.55$\%. Also shown in Fig.~\ref{2D_and_3D_plots}b is the integrated intensity of the $b$ component under compression along the $b$ axis. The effect of $b$ axis compression is opposite to that of $a$ axis compression: the intensity shrinks. In Supplementary Fig. 5 of Supplementary Note 2 we provide evidence confirming the growth of the $a$ axis component of the 2D CDW under $b$ axis compression, in agreement with previous findings~\cite{Kim2021}. Together, these data show that the $a$- and $b$-CDWs are quasi-independent order parameters, and emphasize the 
uniaxial nature of the underlying order parameter.

The decrease in intensity of the 2D CDW for $\varepsilon_{xx} < -0.55$\% coincides with the onset of the 3D CDW. The 3D CDW manifests itself as a sharp peak at integer $L$, signaling long-range, in-phase correlations along the $c$ axis~\cite{Gerber2015, Chang2016}; the 3D CDW intensity is also shown in Fig.~\ref{2D_and_3D_plots}b. It is worth noting that this onset strain is considerably lower than that estimated in previous x-ray scattering studies~\cite{Kim2018, Kim2021}, in which the lattice parameter longitudinal to the direction of compression was not accessible, but is closer to the location of an anomaly identified in the stress dependence of $T_\text{c}$ in Ref.~\cite{Barber2022}. The 3D CDW peak is also seen for $y = 6.55$ ($p = 0.108$) (see Fig.~\ref{discommensuration}b), but not yet for $y = 6.80$ ($p=0.140$), where, admittedly, the largest strain achieved in the present work was lower: we reached $\varepsilon_{xx} = -0.45$\%. (See Supplementary Fig. 6 in Supplementary Note 2.)

We turn next to the temperature dependence of the 2D and 3D CDW intensities. Integrated intensities of the 2D and 3D CDWs versus temperature at selected, fixed strains are shown in Fig.~\ref{temperature dependence}a for YBCO$_{6.67}$, and Fig.~\ref{temperature dependence}b for YBCO$_{6.55}$. At both of the selected strains for both compositions, the 2D CDW intensity stops increasing and (within experimental accuracy even seems to decrease) when the 3D CDW onsets, as was found above to occur for the strain scan on YBCO$_{6.67}$. At these selected strains, the onset temperature of the 3D CDW exceeds $T_\text{c}(\varepsilon_{xx}=0)$ and so, presuming that $a$ axis compression suppresses $T_\text{c}$ in YBCO$_{6.55}$ as it does in YBCO$_{6.67}$ (and as inferred from thermal expansion in the zero stress limit~\cite{Kraut1993}), it substantially exceeds $T_\text{c}$ at these strains. Therefore, the end of the 2D CDW growth at these strains is clearly tied to the onset of the 3D CDW, and not to the onset of superconductivity as at $\varepsilon_{xx} \sim 0$. Intensity maps of the 3D CDW in strain-temperature space, built up from temperature scans over these and additional strains, are shown for YBCO$_{6.67}$ and YBCO$_{6.55}$ in Fig.~\ref{temperature dependence}c and d, respectively. For YBCO$_{6.55}$ there is no direct measurement of $T_\text{c}(\varepsilon_{xx})$, so within the 3D CDW phase we estimate this quantity from the temperature where the 3D CDW signal intensity reaches a maximum, before getting gradually suppressed. This indicates a substantial range of overlap between the 3D CDW and superconductivity. It is on the one hand consistent with measurements in high magnetic fields that systematically find the 3D CDW onset field to lie below $H_{c2}$~\cite{Chang2016} (at least of its lower estimated bound), and on the other hand the 3D CDW signal being observed down to at least 41~K, \textit{i.e.} well below $T_\text{c}(\varepsilon_{xx})$ for YBCO$_{6.67}$ at the highest measured strain of Ref.~\cite{Kim2018} (the present data allow us to re-estimate this value to be $\varepsilon_{xx}=-0.85$\%). For YBCO$_{6.55}$, the 3D CDW onset lies at a slightly smaller strain, $\varepsilon_{xx} = -0.4$\%, lower than for YBCO$_{6.67}$.

Comparing Fig.~\ref{discommensuration}a and b, which show scattering peaks from the 3D CDW at $\varepsilon_{xx} = -0.85$\% in YBCO$_{6.67}$ and YBCO$_{6.55}$, respectively, it can be seen that the peak is substantially narrower for YBCO$_{6.55}$, indicating a longer correlation length, even though the tendency to 3D CDW order is probably stronger in YBCO$_{6.67}$. 
This observation points to a substantial effect of disorder: the $b$ axis correlation length of the ortho-II chains in YBCO$_{6.55}$ is longer than that of the ortho-VIII chains in YBCO$_{6.67}$. 

We end this section with a description of structural changes induced by the 3D CDW formation which can be quantified, despite experimental limitations on the accessible part of the reciprocal space imposed by the strain cell. Within experimental uncertainty, we observe the main structural changes in the CuO$_2$-planes as the samples are strained. From structural refinements of YBCO$_{6.67}$ we observe that the clearest response of the average structure (single unit cell) to the 3D CDW onset with increasing strain is a growing anisotropy of the buckling angle of the planar oxygen bonds as shown in Supplementary Fig. 7 in Supplementary Note 3.  Minor changes can be discerned in the parameters of the planar Cu atoms,
\textit{e.~g.} in the intra-bilayer distance, but these are less evident than the change observed for the buckling angles of the oxygen bonds. Nevertheless, smaller motions are to be expected for the heavier Cu atoms. This also finds confirmation in the smaller isotropic temperature factors, shown in Supplementary Fig. 8 in Supplementary Note 3.

\begin{figure*}[t!!!]
\begin{center}
\begin{tabular}{ll}
\includegraphics[width = 0.492\textwidth]{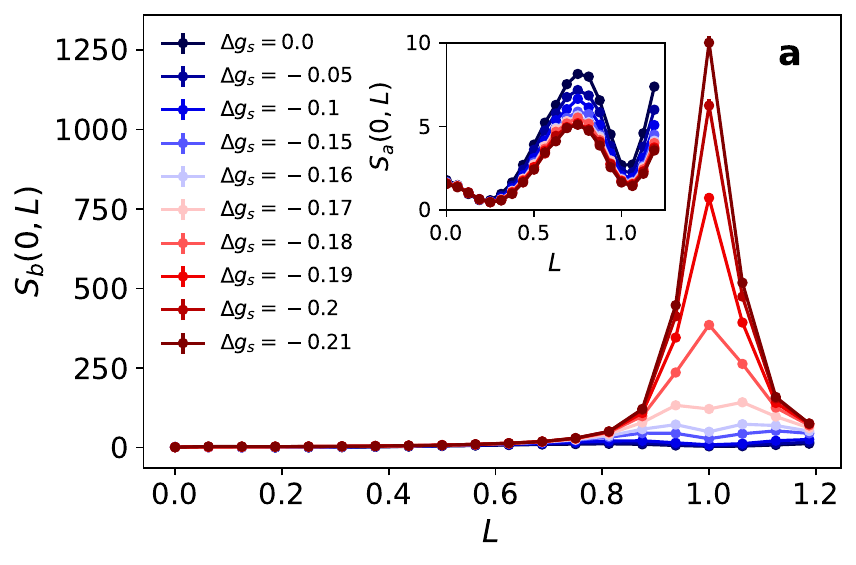}  &
\includegraphics[width = 0.492\textwidth, height = 0.329\textwidth]{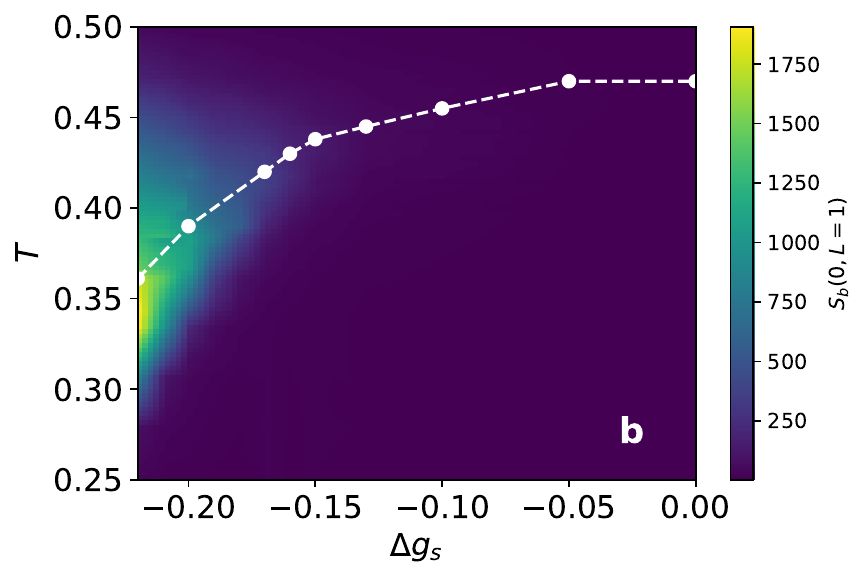} \\
\includegraphics[width = 0.492\textwidth]{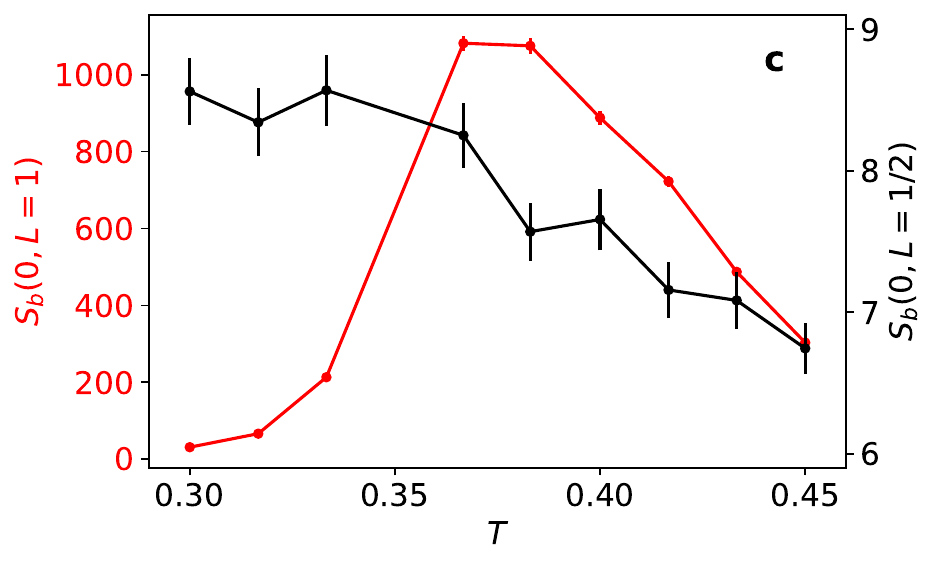} &
\includegraphics[width = 0.492\textwidth]{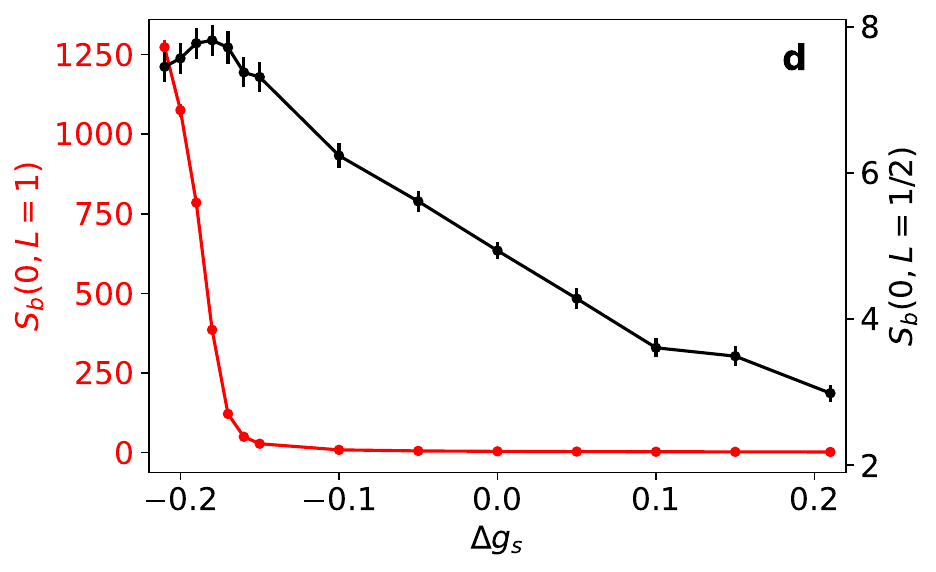}
\end{tabular}
\caption{\textbf{Theoretical CDW structure factors based on a non-linear sigma model.} (a) The CDW structure factor at the in-plane wave vector of the $b$-CDW peaks 
as a function of the $c$ axis wave vector, $L$, for $T=0.383$ and various levels of the change in the CDW-mass, 
$\Delta g_s$, that model the effect of $a$ axis compression. The inset depicts the CDW structure factor at the position  
of the $a$-peak under the same conditions. (b) The intensity of the $L=1$ $b$-CDW peak in the $T$-$\Delta g_s$ plane. 
The broken line marks the superconducting transition temperature. (c) The temperature dependence of the CDW structure 
factor at the position of the $b$-CDW peaks for $L=1$ and $L=1/2$ in a system with $\Delta g_s$=-0.2. (d) The $\Delta g_s$ 
dependence of the same quantities at $T=0.383$.
}
\label{fig:model-res}
\end{center}
\end{figure*}

{\bf Model.} In an effort to elucidate the experimental findings we consider a nonlinear sigma model describing competition
between superconductivity and CDW orders \cite{Hayward_Science2014,Caplan2015,Caplan2017}. It is formulated in terms of a three-dimensional
complex vector whose components correspond to a superconducting order parameter, $\psi_{j\mu}(\br)$, and
two complex CDW order parameters, $\Phi_{j\mu}^{a,b}(\br)$. The latter describe density variations
$\delta\rho_{j\mu}(\br)=e^{i\bQ_a \cdot \br}\Phi_{j\mu}^a(\br)+e^{i\bQ_b \cdot \br}\Phi_{j\mu}^b(\br) +{\rm c.c.}$,
along the $a$ and $b$ directions with incommensurate wave-vectors $\bQ_{a,b}$. The order parameters reside on
bilayers that are indexed by $j$ and represent the CuO$_2$ bilayers of YBCO. The index $\mu=0,1$ denotes the bottom (top) layer.
In the following we will coarse grain each of the planes into a square lattice whose lattice constant is
the observed CDW wavelength, \textit{i.e.}, about 3 Cu-Cu spacings. The model assumes the existence of some type of local
order at every lattice point and incorporates the competition between the different components via the constraints
\begin{equation}
\label{meq:constraint}
|\psi_{j\mu}|^2+|\bPhi_{j\mu}|^2=1,
\end{equation}
where ${\bm \Phi}_{j\mu}=(\Phi_{j\mu}^a,\Phi_{j\mu}^b)^T$.

The vector of order parameters is governed by the Hamiltonian
\begin{eqnarray}
\label{meq:HV}
\nonumber
H&=&\sum_j\sum_{\mu=0,1} H_{j\mu}+\frac{\rhos}{2}\sum_j\sum_\br
\Big[\tU \bPhi^\dagger_{j,0}\bPhi_{j,1}\\
\nonumber
&&+U\bPhi^\dagger_{j,1}\bPhi_{j+1,0}
-\tJ\psi_{j,0}^*\psi_{j,1}-J\psi_{j,1}^*\psi_{j+1,0}\\
\nonumber
&&+{\bm V}^\dagger_j\left(\gamma \bPhi_{j,0}
+ \bPhi_{j,1}+\bPhi_{j+1,0}+\gamma \bPhi_{j+1,1}\right)+{\rm c.c.}\Big]. \\
\end{eqnarray}
Henceforth, the superconducting stiffness, $\rhos$, serves as the basic energy scale.
We model the Coulomb interaction between CDW fields within a bilayer by a local coupling $\tU$, and denote
the intra-bilayer Josephson tunneling amplitude by $\tJ$. The (weaker) Coulomb interaction and Josephson coupling
between nearest-neighbor planes belonging to consecutive bilayers are denoted by $U$ and $J$, respectively.
In the last term, we include the Coulomb interaction between the disordered doped oxygens on the chain layers
and the CDW fields on the adjacent bilayers, assuming that the coupling to the outer CuO$_2$ planes
is reduced by a factor $\gamma$ compared to the coupling to the inner planes. We model the disorder
as a collection of randomly placed discs, each of radius $r_d$ and containing a constant potential with a random phase
that couples either to the $a$ or $b$ CDW fields, ${\bm V}(\br)=V\sum_l f(|\br-\br_l|)e^{i\theta_l}(p_l,1-p_l)^T$.
Here, $p$ takes the values 0,1 with probability 1/2 and $f(\br)=1-\Theta(|\br|-r_d)$, where $\Theta(r)$ is the step function.

Within a layer the Hamiltonian reads
\begin{eqnarray}
\label{meq:slh}
\nonumber
\!\!\!\!\!\!\!H_{j\mu}&=&\frac{\rho_s}{2}\sum_\br\Big[\left|\bnabla\psi_{j\mu}\right|^2 +\lambda|\bnabla \bPhi_{j\mu}|^2
+g|\bPhi_{j\mu}|^2\\
\!\!\!\!\!\!\!&&
+\left(\Delta g -\Delta g_s\right) |\Phi^a_{j\mu}|^2 +\Delta g_s|\Phi^b_{j\mu}|^2 \Big],
\end{eqnarray}
where $\bnabla$ is the discrete gradient, $\lambda\rhos$ is the CDW stiffness and $g\rhos$ is the effective CDW mass reflecting
the energetic penalty for CDW ordering. The presence of such a penalty ensures that superconductivity prevails over the CDW order
at $T=0$, at least in the disorder-free regions. The mass anisotropy, encapsulated by the $\Delta g>0$ term, is included as a result
of our assumption that the potential induced by the chain layers favors ordering along the $b$ axis. Finally, and most pertinently
to the present study, we assume that the application of strain causes an increase in the mass of the CDW component along the
direction of the strain, while reducing the mass of the transverse component. Specifically, applying compressive strain in the $a$ direction
corresponds to $\Delta g_s<0$.

{\bf Theoretical results.}
Our primary interest lies in the $k$-space (measured from $\bQ_{a,b}$) CDW structure factor
\begin{eqnarray}
\label{meq:Sdef}
\nonumber
S_\alpha(\bq,L)&=&\frac{1}{N}\sum_{\br\br'}\sum_{jj'}\sum_{\mu\mu'} e^{-i\left[\bq\cdot \left(\br-\br'\right)
+ 2\pi \left(j-j'+\frac{\mu-\mu'}{3}\right)L\right]}\\
&&\times \langle\Phi^\alpha_{j\mu}(\br)\Phi^{*\alpha}_{j'\mu'}(\br')\rangle,
\end{eqnarray}
where $N$ is the number of lattice points, and the averaging is over both thermal fluctuations and disorder realizations.
We have used the fact that in YBCO the CuO$_2$ planes within a bilayer are separated by approximately 1/3 of the $c$ axis lattice constant.

We have calculated $S_\alpha(\bq,L)$ by Monte Carlo simulations of Eqs. (\ref{meq:HV},\ref{meq:slh}) on a $32\times 32\times 32$
(16 bilayers) system, with $\lambda=1$, $g=1.1$, $\Delta g=0.1$, $\tJ=0.15$, $J=0.015$, $\tU=0.85$,
$U=0.12$, $V=1$, and $\gamma=0.15$. Each data point was averaged over 1000 disorder realizations with 8 disordered regions
per bilayer and $r_d=3$ \cite{disordercomment}.

Our core result is presented in Fig. \ref{fig:model-res}a, which depicts the structure factor at a temperature near
the superconducting $T_c$ and for various values of $\Delta g_s$, emulating the effect of $a$ axis strain. For $\Delta g_s=0$
(absence of strain), both $S_a$ and $S_b$ show broad peaks centered around $L=0.7$. This is a result of CDW domains
that nucleate due to interaction with the disorder \cite{Caplan2017}. Since a disordered region tends to induce the same
CDW pattern on both its flanking planes, an out-of-phase arrangement of the CDW order tends to form in the $c$-direction.
The observed skewness away from $L=1/2$ is a result of the form factor in Eq. (\ref{meq:Sdef}). Increasing $\Delta g_s$ towards
negative values increases the energetic cost for nucleating $a$-CDWs and leads to a decrease in the height of the corresponding
peak (see inset). An opposite trend is seen for the disorder-induced peak along the $b$ direction. However, a much more dramatic
effect emerges in $S_b$ beyond a characteristic value of $\Delta g_s$ in the form of a large sharp peak centered at $L=1$.
This signal originates from the regions {\it between} the disorder-induced CDW domains. It reflects the tipping of the balance between superconductivity and the $b$-CDW order in favor of the latter when the $b$-CDW nucleation cost is sufficiently reduced.
The Coulomb interaction between the strain-induced CDW regions on neighboring bilayers favors an in-phase CDW configuration along 
the $c$-direction, hence the peak at $L=1$. The sharpness of the peak is a direct result of the fact that these regions are not 
induced by disorder.

The interplay between the superconducting and CDW orders is summarized in Fig. \ref{fig:model-res}b, which shows 
the superconducting $T_c$ and the height of the $L=1$ CDW peak as a function of the temperature and $\Delta g_s$. Clearly, 
$T_c$ (calculated by finite size analysis of the superconducting correlation length and stiffness) is a decreasing function 
of $|\Delta g_s|$ owing to the enhancement of the competing CDW fluctuations. A large 3D CDW signal appears for 
$\Delta g_s\lesssim -0.17$ at a temperature that increases with $\Delta g_s$, peaks in the vicinity of $T_c$ 
of the strained system and then rapidly diminishes at lower 
temperatures, as is also evident in Fig. \ref{fig:model-res}c. This behavior stands in contrast to behavior of $S_b(0,L=1/2)$ 
that similarly increases when the temperature is lowered towards $T_c$ but then saturates. 

Finally, Fig. \ref{fig:model-res}d shows the $\Delta g_s$ dependence of the 2D and 3D CDW signals near $T_c$. 
There is a clear and sharp onset of the 3D CDW peak at $\Delta g_s\simeq -0.17$, whereas $S_b(0,L=1)$ essentially 
vanishes at lower values of $a$ axis strain and for all values of $b$ axis strain (positive $\Delta g_s$). 
The 2D signal, $S_b(0,L=1/2)$, on the other hand, diminishes continuously with decreasing $a$ axis strain and increasing $b$ axis strain. 
This is because the accompanying increase in the effective mass of the $b$-CDW successively reduces the magnitude of the 
CDW induced by the disorder. Interestingly, and similarly to the experimental observations, 
we find a saturation and then a slight downturn of the 2D CDW signal that coincides with the onset of the 3D CDW order.

\section{Discussion} 
\label{sec:discussion}
We start our discussion from the structure of the 3D CDW. Having access to large reciprocal space maps and intense 3D CDW peaks, we can provide strong constraints for deviations from a perfectly sinusoidal charge density modulation in the long-range ordered phase. Previous studies on Bi-based cuprates~\cite{Mesaros2016} have indicated that in this system the CDW is commensurate with periodicity $\lambda=4a$ over the entire doping range where it appears. More recently, high-field NMR data have been successfully analyzed in a picture where the CDW is locally commensurate with $\lambda=3b$. In both cases it was argued that sharp phase slips (discommensurations) at the CDW domain boundary could shift the position at which the 3D CDW peak is observed in scattering experiments to an effectively incommensurate value~\cite{Mesaros2016, Vinograd2021}. In this case, one would expect the presence of satellite peaks at $q_{\rm{CDW}}\pm 2\delta$ ($\delta$ being the incommensurability to the commensurate 3D CDW wavevector, $1/3$)~\cite{Vinograd2021}. At $T_c$, for YBCO$_{6.67}$ we observe no indications for additional satellite peaks along $b^{*}$ coming from such sharp discommensuration (we cannot fully rule out the presence of spatially extended phase slips). 
We also note that the $K$-cut shown in Fig.~\ref{discommensuration}b for YBCO$_{6.55}$ at $T=45$~K~$<T_c$ does not exhibit a peak at 
$q_{\rm{PDW}}=q_{\rm{CDW}}/2$. Such a peak is expected if the observed CDW is due to a pair density wave (PDW), \textit{i.e.}, a modulated 
superconducting condensate, at wavevector $q_{\rm{PDW}}$, and if the PDW coexists with uniform superconductivity~\cite{Agterberg2020,Blackburn2023}. 
Based on our measurements, the PDW signal must be at least 50 times smaller than the signal of the long-range 3D CDW. However, further calculations are needed to estimate the associated periodic lattice displacement and the resulting x-ray intensity to exclude the presence of PDWs.

Next, and before discussing the nature of the interplay between the 2D and 3D CDWs, it is worth noting that from a structural point of view, the fact that the formation of the 3D CDW mostly results in a displacement of planar oxygen atoms is consistent with a larger response of the relative quadrupolar splitting in $^{17}$O-NMR than in $^{63}$Cu-NMR when the 3D CDW is induced using high magnetic fields~\cite{Vinograd2021}. Structural refinement of the x-ray diffraction data in zero field shows that the 2D CDWs along $a$ and $b$ axes induce an anisotropic modulation of the copper-oxygen-bonds with significant out-of-plane displacements of the oxygen transverse to the CDW modulation~\cite{Forgan2015}.
In this way the 2D CDW locally breaks rotational symmetry and has a nematic structural character.  
In fact, various experimental probes associate nematicity in the cuprates with the growth of spin and charge density 
wave correlations upon cooling~\cite{Ando2002,Cyr-Choniere2015,Wu2015,Frachet2021}. 
Here, for $a$ axis compression, we observed an increasing trend of the $a$ axis buckling angle close to the 3D CDW
onset, although the error bars are relatively large compared to this subtle effect. In the unstrained orthorhombic unit cell the buckling angles along $a$ and $b$ axes differ only slightly. Hence, the increased strain-induced buckling anisotropy can be regarded as a structural parameter possibly related to nematicity.  Whereas anisotropy in the buckling is expected to exist for 2D CDWs, it is maximal in the 3D CDW phase. However, as the 3D CDW breaks rotational and translational symmetry globally, the anisotropy of the 3D CDW should not be considered as nematic~\cite{Grissonnanche2022}. Further, the buckling anisotropy should not be regarded as an order parameter of the 3D CDW, as for applied strain it is already finite at temperatures higher than that of the 3D CDW formation. We conclude from this that a buckling anistropy might be a necessary condition for 3D ordering in YBCO, but it is certainly not sufficient. In any event, to better understand the constraints imposed by the crystal structures on the formation of long-range charge order in the cuprates (in particular in view of the recent reports on the overdoped side~\cite{Peng_NatMat2018,Tam_NatCom2022}), more systematic structural investigations are needed.\\

We now turn to the nature of the interplay between the CDW orders and superconductivity. 
On the one hand, a large body of experiments has unambiguously established that the 2D CDW and superconducting orders compete in 
YBCO~\cite{Ghiringhelli2012, Chang2012, Blanco2014}, as both the CDW peak amplitude and the CDW domain correlation length decrease 
in the superconducting state. On the other hand, the strongest 3D CDW signals are observed where the homogeneous superconducting 
phase has been suppressed either by application of a large magnetic field~\cite{Choi2020,Jang2016,Chang2016,Gerber2015} or by uniaxial 
pressurization~\cite{Kim2018, Kim2021}. In both cases, the maximal achievable strength of the applied perturbations was not sufficient 
to completely suppress the superconducting state at the lowest temperatures. However, while in the case of strain tuning the low-temperature 
superconducting phase is homogeneous with no trace of the 3D CDW, the mixed state of the type-II superconductor in a magnetic field 
contains halos of 3D CDW around vortex cores.

In the experimental data presented above, we have shown that a form of competition can also manifest itself between the 2D and 3D CDW orders, which is best evidenced in the region of the strain-temperature space where superconductivity has been suppressed by strain.
Data in unstressed YBCO$_{6.67}$ under applied magnetic field~\cite{Choi2020,Jang2016,Chang2016,Gerber2015} can also be
interpreted in terms of this 2D-3D CDW competition. Specifically, it was shown in Ref.~\cite{Choi2020} that as YBCO$_{6.67}$ 
is cooled under strong applied field there is a temperature range where the 2D CDW intensity starts to decrease while the 3D CDW 
keeps increasing.   
In the same study, a direct conversion of the 2D CDW to the 3D CDW was hypothesized, but the data there do not allow clean disentanglement of the mutual interaction among the 2D CDW, 3D CDW, and superconductivity. 
Our data clearly indicate that the 2D CDW stops growing and even shrinks at the onset of the 3D CDW, where the superconductivity would have set in, in the absence of strain. In the superconducting state, the evolution of 2D CDWs is complicated by their mutual competition with the 3D CDW and the simultaneous competition with superconductivity.

Finally, we note that the different temperature- and strain-dependence of the 2D and 3D CDWs can naturally be explained 
in the framework of the phenomenological model developed above and is rooted in 
the different mechanisms that are responsible for their respective establishment. 
The 2D CDW arises in domains that are dominated by their coupling to the disorder potential. Hence, these domains continue 
to host the CDW even when superconductivity sets in within the intervening disorder-free regions. In contrast, the 3D CDW 
arises in the same intervening regions at large enough strain. 
This qualitatively explains why the 2D CDW order is stronger in YBCO$_{6.67}$ than YBCO$_{6.55}$~\cite{Blanco2014} (the former being more disordered than the latter) and why on the contrary the 3D CDW order onsets at a smaller strain in YBCO$_{6.55}$ compared to YBCO$_{6.67}$.
The strain further reduces the energetic cost of the CDW fluctuations, thus allowing them to appear at temperatures that can even exceed the $T_c$ of the unstrained system. 
In turn, the growth of the 3D CDW correlations contribute, via their competition with superconductivity, to the suppression of $T_c$.
Below $T_c$ the 3D CDW eventually gives way to the superconducting order, which constitutes the ground state of the clean model.
Within the same picture, the coincident saturation (or shrinkage) of the 2D CDW with the establishment of strong 3D CDW can be traced to phase mismatch between the two types of CDWs. The phase of the 2D CDW within the disordered domains conforms to the local disorder arrangement, whereas the 3D CDW seeks to establish a uniform phase throughout the system. The phase mismatch along the boundaries of the disordered domains gives rise to an effective surface tension, via the CDW elastic term in Eq. (\ref{meq:slh}), that consequently arrests the growth of the 2D CDW domains.\\
These results more generally highlight the inherent strengths of strain tuning as a powerful approach to investigate and manipulate the intricate interplay between competing orders in quantum materials, presenting promising prospects for advancing our understanding of these systems.

\section*{Methods}
{\bf Samples.}
High-quality single crystals of \ybco~with oxygen concentrations $6.55\leq \rm{y} \leq6.80$ were grown, detwinned and characterized as previously described in ref.~\cite{Blanco2014}. The crystals were then cut and polished to needles with dimensions of $\approx 200~\mu$m$\times100~\mu$m$\times$2~mm, and were pressurized along their lengths using a Razorbill CS200T piezoelectric-driven uniaxial stress cell mounted in a helium flow cryostat. Information about the samples is summarized in Supplementary Table~I in Supplementary Note 1. To enable fast sample exchange, the mounting into the stress cell was done via a flexible titanium support, as displayed in Fig.~\ref{strain_cell_0KL_6p67}; needles were mounted into these supports in advance, a slow, delicate process, then the supports were mounted into the cell during the beamtime. To achieve higher stresses, the selected samples were further thinned down laterally using a Xenon plasma focused ion beam (PFIB), as described in Supplementary Note 1. We checked that this procedure did not alter their superconducting properties. 

{\bf X-ray Diffraction.}
Hard x-ray diffraction was performed at the ID15B beamline of the European Synchrotron Radiation Facility with a fixed photon energy of 30 keV ($\lambda$=0.413 $\angstrom$), that enabled us to work in transmission geometry. The scattered photons were collected using a state-of-the-art large area Dectris Eiger2X CdTe 9M hybrid photon-counting detector. Additional experimental details are given in the Supplementary Note 1 and the original data are available under~\cite{ESRF_Dataset2}.
Rigaku's CrysalisPro software~\cite{CrysalysPro} has been used for data reduction and to obtain reciprocal space maps. Structural refinements of the average unit cell (neglecting oxygen ordering in the chain layer) have been performed using Jana2020~\cite{Petricek2014}. In spite of the large dynamical range of the detector, the very high intensity difference between the CDW features and the main Bragg reflections presents a measurement challenge. To map both features, photon flux was adjusted by varying the undulator gap~\cite{Kunz2001}: high photon flux was used to study CDW superstructure peaks, while low photon flux was used to record the lattice Bragg peaks without saturating the detector. Additional details about the structure refinements are given in the Supplementary Note 3.

{\bf Strain determination.}
Strains reported in this paper are derived from the Bragg peak positions, eliminating error from uncertainty in strain transmission to the sample. The determination of lattice parameters was based on integration of about 400 Bragg reflections, a number that was limited by the geometrical constraints imposed by the cryostat and the stress cell, but which is nevertheless sufficient to allow accurate determination of the three lattice constants $a$, $b$, and $c$. Therefore, we also obtain the Poisson's ratios, $\nu_{ij} \equiv-\varepsilon_{ii}/\varepsilon_{jj}$, shown in Supplementary Fig. 2 in Supplementary Note 1. Knowledge of the Poisson's ratios is useful both as a reference for other measurements, where the scattering geometry might not allow determination of the longitudinal strain~\cite{Kim2018}, and potentially as a thermodynamic probe of possible changes in the electronic structure~\cite{Noad2023}. Note that throughout this paper we use the engineering definition of strain, in which negative values denote compression.

\section*{Acknowledgements}
We thank M. Du\v{s}ek, M. Hanfland, S. Kivelson, V. Pet\v{r}\'{i}\v{c}ek, J. Schmalian and R. Willa for helpful discussions, A. K. Jaiswal for SQUID measurements, A. Ghiami and M. Hesselschwerdt for technical support and T. Poreba and N. Maraytta for support during the diffraction experiments at the ESRF and IQMT, respectively. Self-flux growth was performed by the Scientific Facility ‘Crystal Growth’ at Max Planck Institute for Solid State Research, Stuttgart, Germany. This work was supported through the funding of the Deutsche Forschungsgemeinschaft (DFG, German Research Foundation), projects 422213477 (TRR 288 projects B03 and A10) and 449386310. S.M.S. acknowledges funding by the DFG – Projektnummer 441231589, M. F. acknowledges funding by the Alexander von Humboldt Foundation and the Young Investigator Grant, H. M. L. N. acknowledges support from the Alexander von Humboldt Foundation through a Research Fellowship for Postdoctoral Researchers, K. I. acknowledges the Japan Society for the Promotion of Science Overseas Research Fellowships and I. V. acknowledges the Horizon Europe MSCA fellowship 101065694. We thank the European Synchrotron Radiation Facility (ESRF) for provision of synchrotron radiation facilities under proposals number HC-4226 and HC-4865. 

\noindent\textbf{Data availability}
The data reported in this study are available at \cite{ESRF_Dataset2}. 
The data that support the findings of this study are available from the corresponding author,  M.~L.~T.(\href{mailto:matthieu.letacon@kit.edu}{matthieu.letacon@kit.edu}), upon reasonable request.

\bibliography{strainbib}

\begin{thebibliography}{55}
\bibitem [{dis()}]{disordercomment}%
  \BibitemOpen
  \href@noop {} {\bibinfo {title} {{The disorder potential in an overlap region between two discs with different disorder orientations was taken to be the sum of the potentials. In the case of two discs with the same orientation the potential in the overlap region was randomly chosen to be one of the two.}}}\BibitemShut {Stop}%
\end{thebibliography}

\end{document}


\title{Supplementary Information\\Using strain to uncover the interplay between tow- and three-dimensional charge density waves in high-temperature superconducting \ybco}

\author{I. Vinograd}
\altaffiliation{These authors contributed equally to this work}
\affiliation{Institute for Quantum Materials and Technologies, Karlsruhe Institute of Technology, Kaiserstr. 12, D-76131 Karlsruhe, Germany}
\affiliation{4th Physical Institute – Solids and Nanostructures, University of Göttingen, D-37077 Göttingen, Germany}
\author{S. M. Souliou}
\altaffiliation{These authors contributed equally to this work}
\affiliation{Institute for Quantum Materials and Technologies, Karlsruhe Institute of Technology, Kaiserstr. 12, D-76131 Karlsruhe, Germany}
\author{A.-A. Haghighirad}
\affiliation{Institute for Quantum Materials and Technologies, Karlsruhe Institute of Technology, Kaiserstr. 12, D-76131 Karlsruhe, Germany}
\author{T. Lacmann}
\affiliation{Institute for Quantum Materials and Technologies, Karlsruhe Institute of Technology, Kaiserstr. 12, D-76131 Karlsruhe, Germany}
\author{Y. Caplan}
\affiliation{Racah Institute of Physics, The Hebrew University, Jerusalem 91904, Israel}
\author{M. Frachet}
\affiliation{Institute for Quantum Materials and Technologies, Karlsruhe Institute of Technology, Kaiserstr. 12, D-76131 Karlsruhe, Germany}
\author{M. Merz}
\affiliation{Institute for Quantum Materials and Technologies, Karlsruhe Institute of Technology, Kaiserstr. 12, D-76131 Karlsruhe, Germany}
\affiliation{Karlsruhe Nano Micro Facility (KNMFi), Karlsruhe Institute of Technology, Kaiserstr. 12, D-76131 Karlsruhe, Germany}
\author{G. Garbarino}
\affiliation{ESRF, The European Synchrotron, 71, avenue des Martyrs, CS 40220 F-38043 Grenoble Cedex 9}
\author{Y. Liu}
\affiliation{Max Planck Institute for Solid State Research, Heisenbergstraße 1, D-70569 Stuttgart, Germany}
\author{S. Nakata}
\affiliation{Max Planck Institute for Solid State Research, Heisenbergstraße 1, D-70569 Stuttgart, Germany}
\author{K. Ishida}
\altaffiliation{Current address: Institute for Materials Research, Tohoku University, Sendai 980-8577, Japan}
\affiliation{Max Planck Institute for Chemical Physics of Solids, N\"{o}thnitzer Str. 40, D-01187 Dresden, Germany}
\author{H. M. L. Noad}
\affiliation{Max Planck Institute for Chemical Physics of Solids, N\"{o}thnitzer Str. 40, D-01187 Dresden, Germany}
\author{M. Minola}
\affiliation{Max Planck Institute for Solid State Research, Heisenbergstraße 1, D-70569 Stuttgart, Germany}
\author{B. Keimer}
\affiliation{Max Planck Institute for Solid State Research, Heisenbergstraße 1, D-70569 Stuttgart, Germany}
\author{D. Orgad}
\affiliation{Racah Institute of Physics, The Hebrew University, Jerusalem 91904, Israel}
\author{C. W. Hicks}
\affiliation{Max Planck Institute for Chemical Physics of Solids, N\"{o}thnitzer Str. 40, D-01187 Dresden, Germany}
\affiliation{School of Physics and Astronomy, University of Birmingham, Birmingham, B15 2TT, UK}
\author{M. Le Tacon}
\email{matthieu.letacon@kit.edu}
\affiliation{Institute for Quantum Materials and Technologies, Karlsruhe Institute of Technology, Kaiserstr. 12, D-76131 Karlsruhe, Germany}

\maketitle

\subsection{Experimental details}
The single crystal samples were grown by a flux technique described in Ref.~\cite{Lin1992} and the sample information is summarized in Supplementary Table \ref{tab:sample info}. Superconducting $T_c$ and $c$ axis parameters were used to determine the hole doping.
The samples were glued into a titanium cross carrier using Stycast FT-2850. A frame was screwed onto the Razorbill CS200T strain cell into which the cross itself was glued. This allows us to perform the delicate step of gluing the sample onto the cross prior to the experiments. The nominal stress application is monitored by a capacitive sensor integrated in the cell. The capacitance is monitored with an AH2550A high-resolution capacitance bridge or a Keysight 4980AL LCR meter and converted into a nominal displacement and strain. The effective strain transmission is estimated from the slope in Supplementary Fig.~\ref{strain_transmission}. Reducing the sample width using a Xenon plasma focused ion beam (Xe-plasma FIB) improves the strain transmission significantly, but the maximally achieved real strain remains just above -1 \%, a value beyond which samples typically break. We confirmed by SQUID measurements that the FIB processing does not modify the sample's bulk $T_c$.   

\begin{figure*}
\hspace*{-0.3cm}    
  \includegraphics[width=15cm]{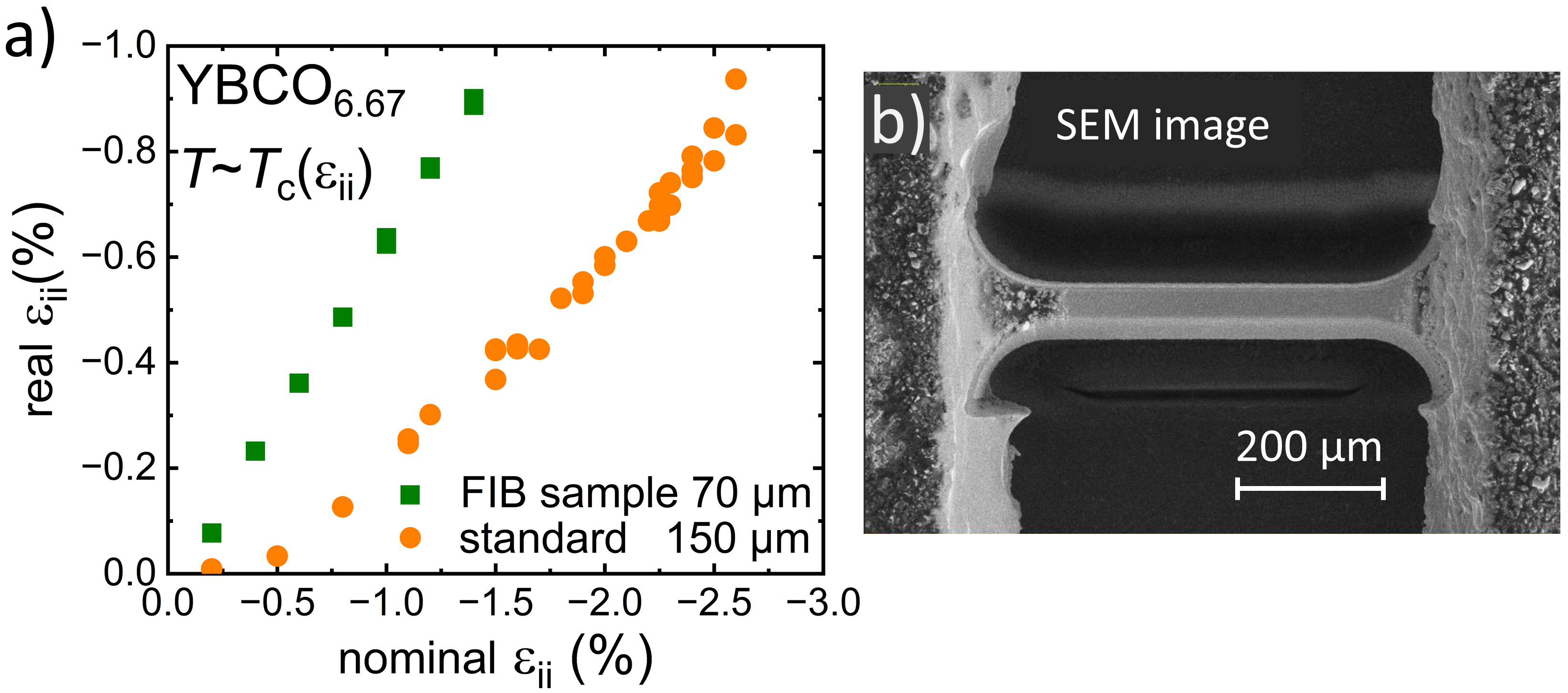}
  \caption{\label{strain_transmission} (a) Strain transmission is estimated from real versus nominal strain for a standard sample (orange circles, $a$-compression) and a sample thinned to 70 $\mu$m using a Xe-plasma FIB (green squares, $b$-compression). The FIB sample displays a better strain transmission. (b) SEM image of the thinned sample after the FIB processing.}
\end{figure*}

\begin{table}[b]
\caption{\label{tab:sample info}%
Sample information and sample geometry (compression axis and crystal axis parallel to beam during transmission at zero rotation)
}
\begin{ruledtabular}
\begin{tabular}{cccc} 
\textrm{oxygen content y} &\textrm{doping $p$}&
\textrm{$T_{\textrm c}$ (K)} &compression \& transmission axis \\

\colrule
 6.67 & 0.125 &  67  & $a$ \& $b$ axis  \\    
 6.67 & 0.125 &  67  & $b$ \& $a$ axis  \\  
 6.55 & 0.108 &  61  & $a$ \& $c$ axis \\  
 6.67 & 0.125 &  67  & $b$ \& $c$ axis \footnote{cut with FIB}\\
 6.80 & 0.140 &  78  & $a$ \& $c$ axis \\
 
\end{tabular}
\end{ruledtabular}
\end{table}
The high-pressure beamline ID15B at ESRF operates with monochromatized 30 keV photons ($\lambda$=0.4101 $\angstrom$) produced by an undulator~\cite{Kunz2001}. The flux is reduced by increasing the undulator gap which shifts the harmonics energy relative to the monochromator. This allows to prevent beam-heating of the sample in high flux measurements and not to over-saturate the lattice Bragg peaks in low flux measurements.
A side-effect of the increased undulator gap is the relative increase in intensity of undesired harmonics which pass the monochromator at $\lambda/2$. This leads to weaker but visible lattice Bragg peaks appearing at half-integer reciprocal lattice values. These $\lambda/2$ peaks are too weak to be visible in the lowest flux data (undulator gap = 15.6-16.6 mm). Higher flux (undulator gap = 13.5 mm) saturates the main lattice Bragg reflections in all presented reciprocal space maps. The sharp $\lambda/2$ peaks can be easily distinguished from CDW peaks from their temperature and strain dependence. The sharpness of the $\lambda/2$ peaks in $(0KL)$ and $(H0L)$ maps are evidence of the high crystal quality, even when the main lattice peaks are over-saturated and appear very broad.  

Since we observe a linear expansion of the transverse crystallographic axes in Supplementary Fig.~\ref{lattice_parameters_6p67}, described by slope $m$ and intercept $b$, we can assume a strain-independent Poisson ratio and calculate the Poisson ratio as $\nu_{ij}=-m/b\times 100\%$. The systematic uncertainty in the determination of $\varepsilon_{ii}=0$ due to differences in the thermal expansions of the strain cell and the sample after cooling down is estimated to lie within $\pm 0.05\%$, as witnessed from the intercept of linear fits for $a$ and $b$ axis compression in Supplementary Fig.~\ref{lattice_parameters_6p67} being close to $\varepsilon_{ii}=0$. That $a$ and $b$ axis compression lead to $\nu_{yx}\neq \nu_{xy}$, for the YBCO$_{6.67}$ samples in Supplementary Fig.~\ref{Poisson_ratio} is expected for the orthorhombic lattice. It means that the Young's moduli, $E_i$ are different, following $\frac{\nu_{yx}}{E_y} =\frac{\nu_{xy}}{E_x}$, however, $\sim$47\% difference is remarkable.
\begin{figure}
\hspace*{-0.1cm}
  \includegraphics[width=8.4cm]{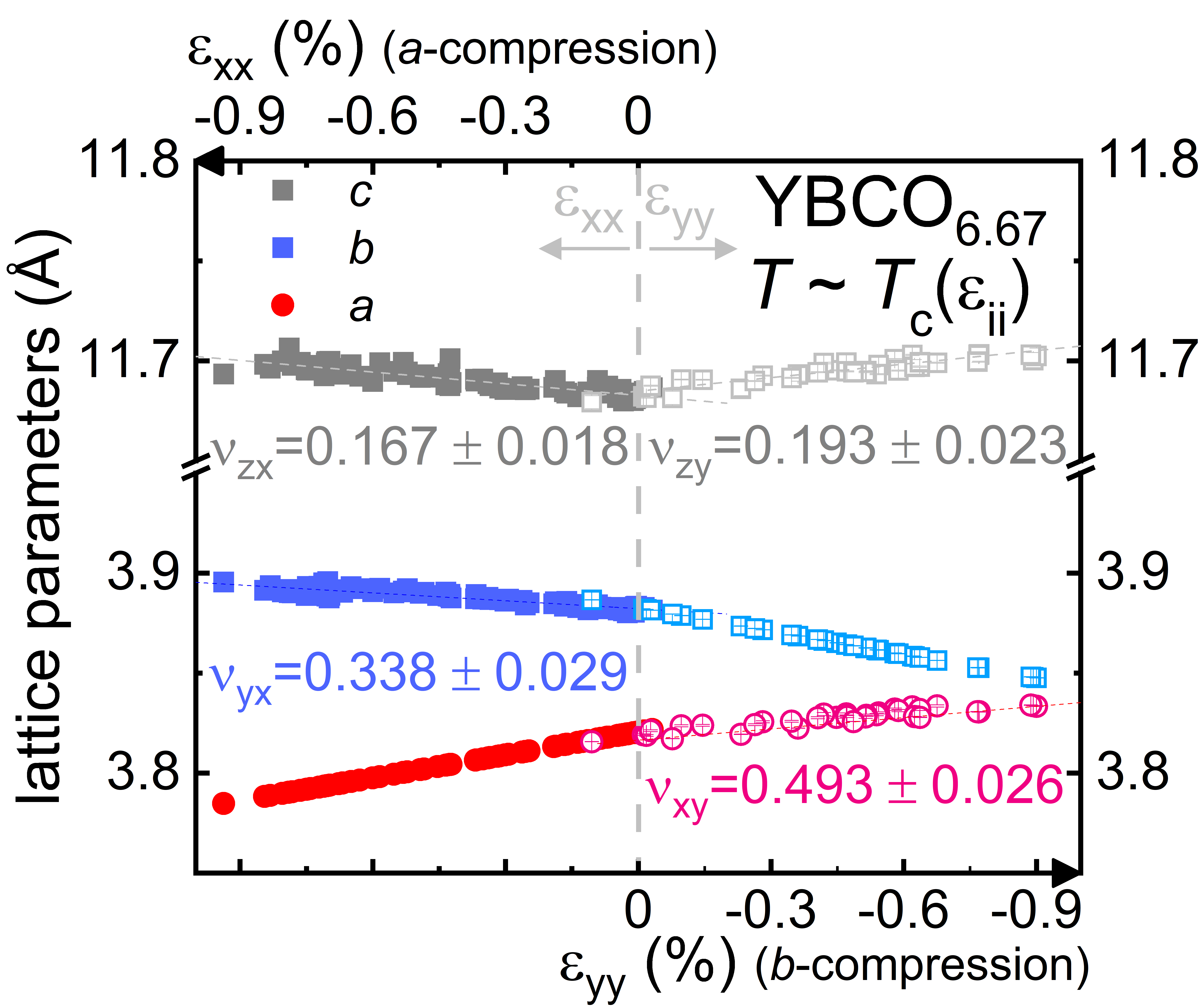}
  \caption{\label{lattice_parameters_6p67} Strain evolution of the unit cell parameters for $a$ and $b$ axis compressions of the three investigated YBCO$_{6.67}$ samples. For linearly expanding transverse directions the Poisson ratios are determined from the linear intercept and the slope.}\end{figure}

\begin{figure}
\hspace*{-0.5cm}    
  \includegraphics[width=6.5cm]{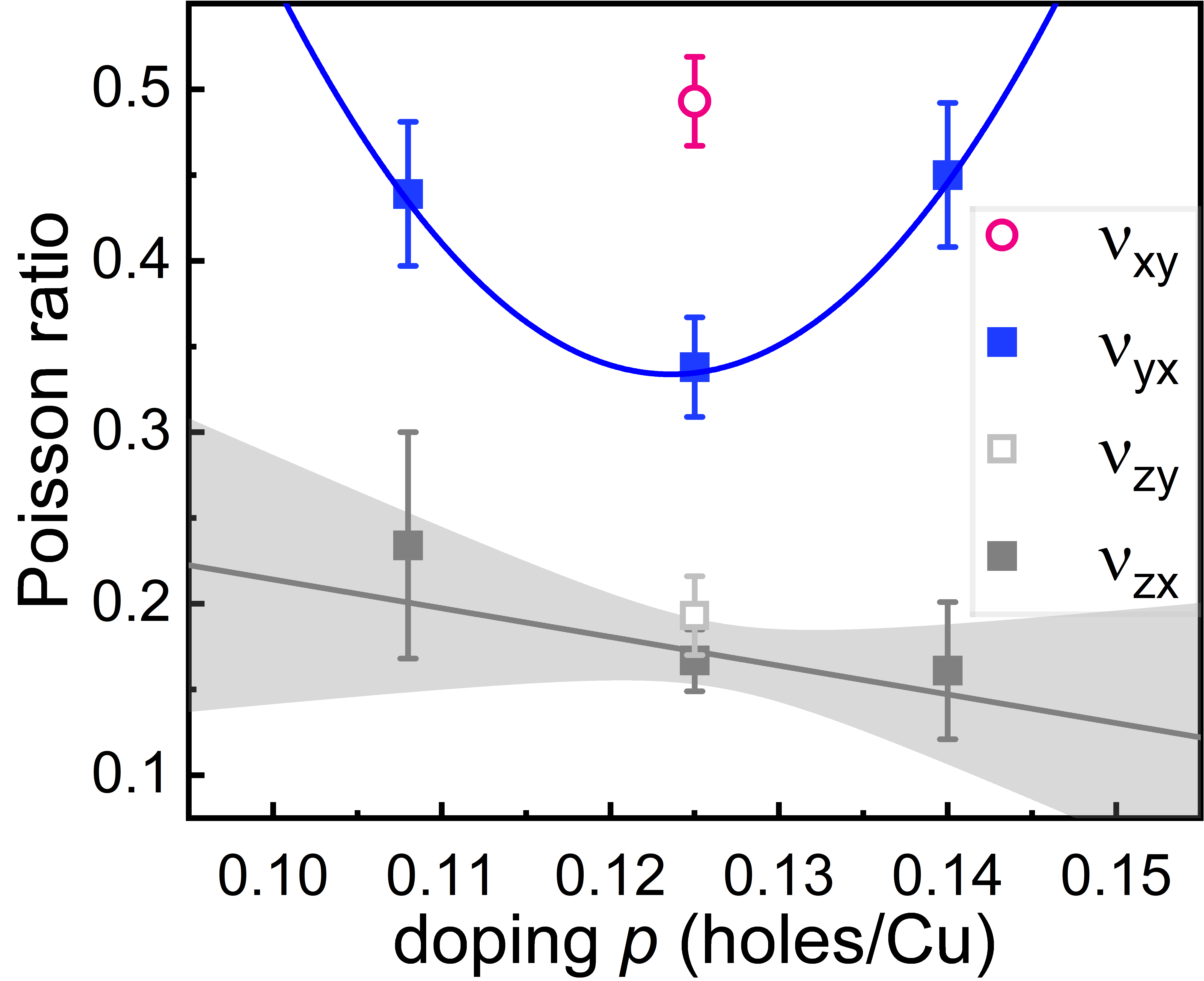}
  \caption{\label{Poisson_ratio} Hole doping dependence of the Poisson ratios for the investigated samples with one-$\sigma$ confidence intervals for the linear fit of $\nu_{zx}$ indicated by shaded grey regions. This one $\sigma$-confidence interval of the fit indicates that $\nu_{zx}$ can be assumed as doping independent up to the experimental precision. The parabola is a guide to the eye for $\nu_{yx}$. Empty symbols correspond to $\nu_{xy}$ and $\nu_{zy}$ determined from $b$ axis compression.}
\end{figure}

\subsection{2D and 3D CDWs}
\begin{figure}
\hspace*{-0.3cm}    
  \includegraphics[width=8.5cm]{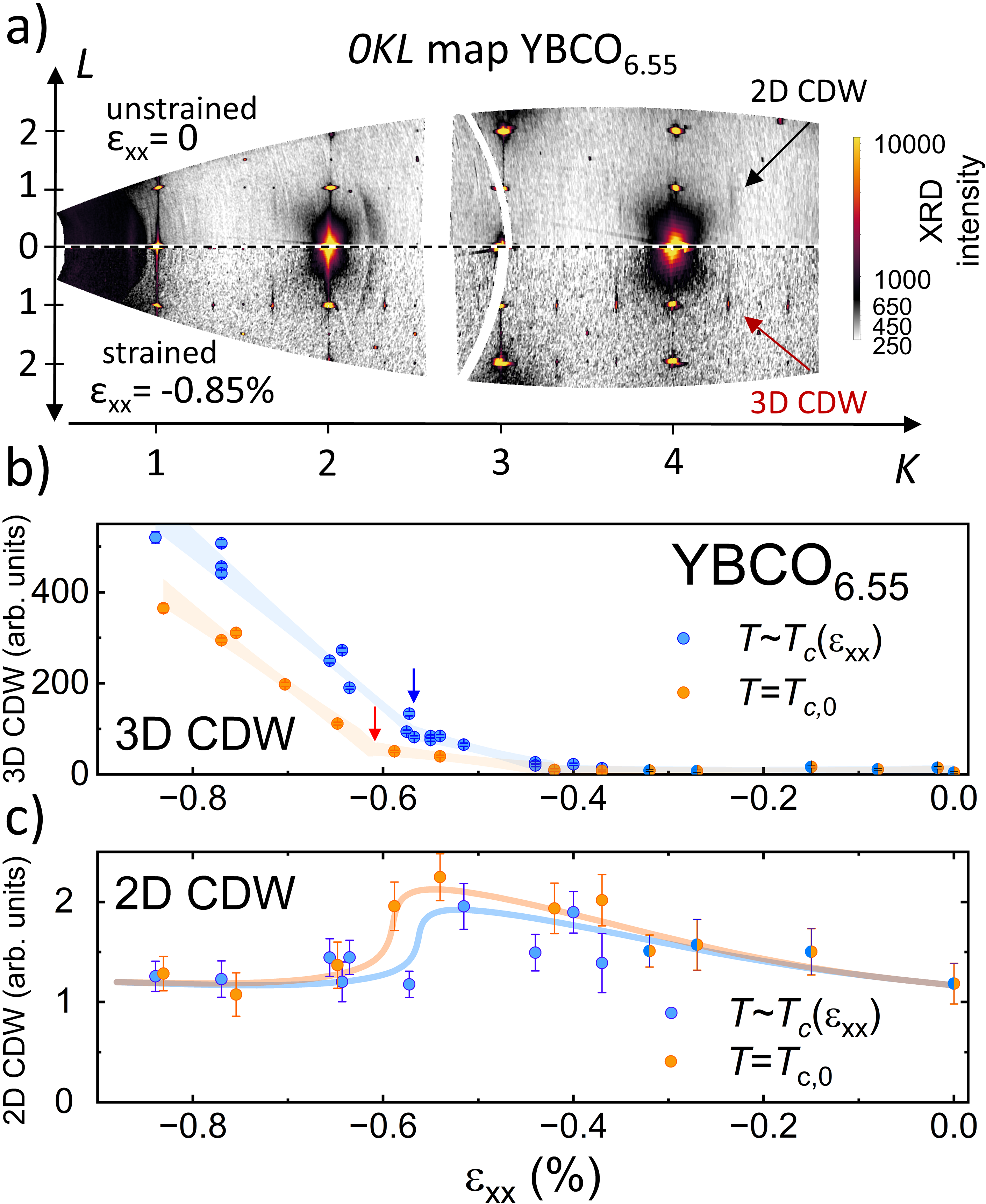}
  \caption{\label{2D_3D_6p55} (a) YBCO$_{6.55}$ ($0KL$) reciprocal space maps at zero (top) and highest $a$ axis compression (bottom) measured at $T_{c,0}$=61~K, show the appearance of sharp 3D CDW peaks at integer $L$ at high compression. Faint 2D CDW intensity in between is present at all strains close to half-integer $L$. (b) Strain dependence of the integrated 3D $b$-CDW peak at $(K,L)=(4+q_{\rm{CDW}},-1)$ from $L$-cuts is maximal at $T\sim T_c (\varepsilon_{xx})$ (blue circles). For comparison, the 3D CDW intensity is shown at constant temperature of $T_{c,0}$=61~K. Lines are piece-wise linear fits (widths correspond to one-$\sigma$ confidence intervals), showing the onset of 3D intensity at strains as low as $\varepsilon_{xx}$=-0.4\% and arrows mark an inflection point, beyond which 3D order grows more rapidly. (c) Strain dependence of the integrated 2D $b$-CDW peak at $(K,L)=(4-q_{\rm{CDW}},0.5)$ from $K$-cuts shows decreasing 2D CDW intensity beyond the strain value where 3D CDW order grows rapidly. Lines are guides to the eye.}
\end{figure}
 Our study mainly focuses on the 2D vs. 3D CDW competition in YBCO$_{6.67}$, as at this doping the 2D CDW is best developed and was consequently studied extensively in previous x-ray diffraction studies~\cite{Chang2012,Chang2016,Choi2020}. In addition, our measurements of the YBCO$_{6.55}$ sample, are consistent with the scenario of direct competition between 2D and 3D CDWs, as shown in Supplementary Figs.~\ref{2D_3D_6p55}b,c. Similarly to YBCO$_{6.67}$, the 2D and 3D CDW intensities increase for $a$ axis compression, but the 2D CDW intensity does not show a cusp at the 3D CDW onset. Only after further compression where the 3D intensity starts to grow more rapidly, as can be seen by the increased linear slope of the growing intensity, the 2D CDW intensity drops. This behavior is illustrated for two datasets: i) The temperatures for which the 3D CDW intensity is maximal, thus likely being close to $T_c(\varepsilon_{xx})$, which decreases with $a$ axis compression. ii) Fixed temperature at $T_{c,0}$=61~K of the unstrained sample. It is interesting to note that the 3D CDW is already well developed in the intermediate strain range $-0.55\% \leq \varepsilon_{xx} \leq -0.40\%$ and the maximal peak intensity of the 3D CDW exceeds that of the broad 2D CDW multiple times. Whereas the peak intensity of the 2D CDW, $I_{max}^{2D}$ is of the order of 100 counts in our measurements for YBCO$_{6.55}$ and YBCO$_{6.67}$, the peak intensity of the 3D CDW, $I_{max}^{3D}$, is $\sim$ 500 counts in the intermediate strain range. The resulting relative 3D peak intensity $I_{max}^{3D}/I_{max}^{2D}\sim$~5 is thus comparable to the highest relative peak intensities reached in high fields up to 26~T~\cite{Choi2020,Gerber2015}. Based on high field NMR measurements up to 45~T, the 3D CDW is already nearly fully developed around $\sim$ 25~T~\cite{Wu2013}. This implies that high magnetic fields alone are not able to reach a 3D CDW state of equal magnitude as by applying large $a$ axis compression. This is likely the reason why high field experiments did not observe this fierce competition of 2D and 3D CDWs, although a plateau in the field dependence of 2D $b$-CDWs is suggestive of a related but weaker effect~\cite{Choi2020}. \\   
 How can the faint 2D CDW's impact on the sharp 3D CDW be explained? Whereas the peak intensities of the 2D and 3D CDWs vary dramatically, the total integrated intensities of the two CDWs are rather comparable, within the range of investigated $a$ axis compression. To estimate the total integrated intensities the correlation lengths $\xi_i=1/\sigma_i$ ($\sigma_i$ is the standard deviation of a Gaussian in the direction $i$) need to be known~\cite{Nakata2022}. Assuming a purely Gaussian peak shape in all three directions of the reciprocal space, the total integrated intensity, $I_{tot}$, is fully determined by the volume fraction occupied by CDWs, $v_{frac}$, the maximal peak intensity $I_{max}$ and the inverse correlation lengths $\xi^{-1}_H$, $\xi^{-1}_K$ and $\xi^{-1}_L$, as
 \begin{equation}
 I_{tot} \propto v_{frac}^{-1}\cdot I_{max} \cdot \xi^{-1}_H \cdot \xi^{-1}_K \cdot \xi^{-1}_L \text{.}   
 \end{equation} 
 These values are listed for the YBCO$_{6.67}$ and YBCO$_{6.55}$ samples in Supplementary Table~\ref{tab:correlation length} and can be compared to the correlation lengths of the O-VIII and O-II chain order, as defects of the chains are thought to pin 2D CDWs~\cite{Caplan2017}. While these correlation lengths are the relevant values to estimate $I_{tot}$ in our experiment, they underestimate the correlation length of the 3D CDW order due to instrumental resolution.   
\begin{table}[b]
\caption{\label{tab:correlation length}%
Correlation lengths, $\xi_i$ of CDW peaks and chains of the YBCO$_{6.67}$ and YBCO$_{6.55}$ samples, as well as highest achieved ($a$- or $b$-compression) $I_{max}$ and corresponding $I_{tot}$, assuming a CDW volume fraction $v_{frac}=1$. Due to different experimental geometry (see Supplementary Table~\ref{tab:sample info}), the largest accessible CDW peaks are $\sim$3 times weaker for the YBCO$_{6.55}$ sample, as estimated from YBCO$_{6.67}$ at zero strain and transmission through the $a$ axis at zero rotation (large ($0KL$) map).}
 \begin{ruledtabular}
\begin{tabular}{cccccc}
\textrm{peak} &$\xi_H$ ($\angstrom$) &$\xi_K$ ($\angstrom$) &$\xi_L$ ($\angstrom$)&
\textrm{$I_{\textit{max}}$ }  
 &  \textrm{$I_{\textit{tot}}$ } \\
\colrule
  $b$-CDW$^{3D}_{6.67}$ & 143 &  97  & 88 & 10000 & 5.8\,10$^{-3}$\\
 $b$-CDW$^{2D}_{6.67}$ & 29 &  73  & 8 & 110 & 4.2\,10$^{-3}$\\    
 $a$-CDW$^{2D}_{6.67}$ & 57 &  58  & 8 & 150 & 3.6\,10$^{-3}$\\ 
 chains$_{6.67}$ &  22 & 73  & 8 &  & \\  
   $b$-CDW$^{3D}_{6.55}$ & 275 &  274  & 84 & 10000 & 1.1\,10$^{-3}$\\
   $b$-CDW$^{2D}_{6.55}$ & 40 &  75  & 8 & 116 & 3.1\,10$^{-3}$\\
 chains$_{6.55}$ &  80 & 182  & 47 &  & \\ 
\end{tabular}
\end{ruledtabular}
\end{table}
From experimental data alone, it is difficult to determine whether the CDW's periodic lattice displacement or the volume fraction is responsible for the increasing CDW peak intensities. However, in the case of 3D CDWs, the correlation lengths are so large that the volume fraction is likely to be close to 1, meaning that $I_{tot}$ mainly grows due to an increasing periodic lattice displacement (squared)~\cite{Forgan2015}. To effectively compete with the 3D CDW, the volume fraction of the 2D $b$-CDW must be large, too. In the theoretical model, its volume fraction is mainly determined by disordered regions due to pinning and is consequently weakly dependent on strain. The 2D CDW peak widths are not changing much with increasing strain, despite the growth in intensity. Hence it is not important whether $H$-, $K$- or $L$-cuts are used to plot 2D or 3D integrated intensities and one can assume that the 2D CDW's intensity is a measure of the periodic lattice displacement, with the correlation lengths $\xi_i$ being approximately limited by the correlation length of the chain order. Given the large widths of the 2D CDW peaks, large strains are needed for the 3D CDW's periodic lattice displacement to surpass its magnitude for the 2D CDW as measured by $I_{tot}$~\cite{Nakata2022}. The apparent competition bears evidence that, in some sense, the faint 2D CDW is on a par with the well-developed 3D CDW order. Consequently, this substantiates the notion that 2D CDWs are also strong enough to compete with and can impact the strength of superconductivity itself for dopings close to $p\sim 0.12$, where $T_c$ and especially $H_{c2}$ become anomalously low~\cite{Grissonnanche2014,Zhou2017}. Thus, the competition between charge density waves and superconductivity is not one-sided and dominated by superconductivity.\\

For the YBCO$_{6.55}$ sample we noted anomalous line shapes of the 3D CDW peaks in $L$-cuts for which the resolution is worse than for in-plane directions. This could be an artifact of the data processing of CrysalisPro, however we cannot exclude occasional saturation of the detector given the sharpness of the 3D peaks. Still, no anomalies are seen in the line shape shown in Fig. 3 b) of the main manuscript. For unstrained or weakly compressed samples we systematically observe very large secondary extinction factors $\sim 1$ in refinements with Jana2020. Despite the low flux used, for certain reflections a few pixels of the detector appear to be saturated and this effect can result in an artificially increased extinction coefficient. For higher compression the lattice peaks broaden slightly and the extinction drops to smaller values. 
At high flux, the 2D CDW intensity is too weak to saturate the detector, yet measurements of the 2D $a$-CDW for the YBCO$_{6.67}$ sample are complicated by effects that could be related to extinction. The 2D $a$-CDW for the YBCO$_{6.55}$ sample was too weak to be studied reliably.  
\begin{figure}
\hspace*{-0.3cm}    
  \includegraphics[width=8.5cm]{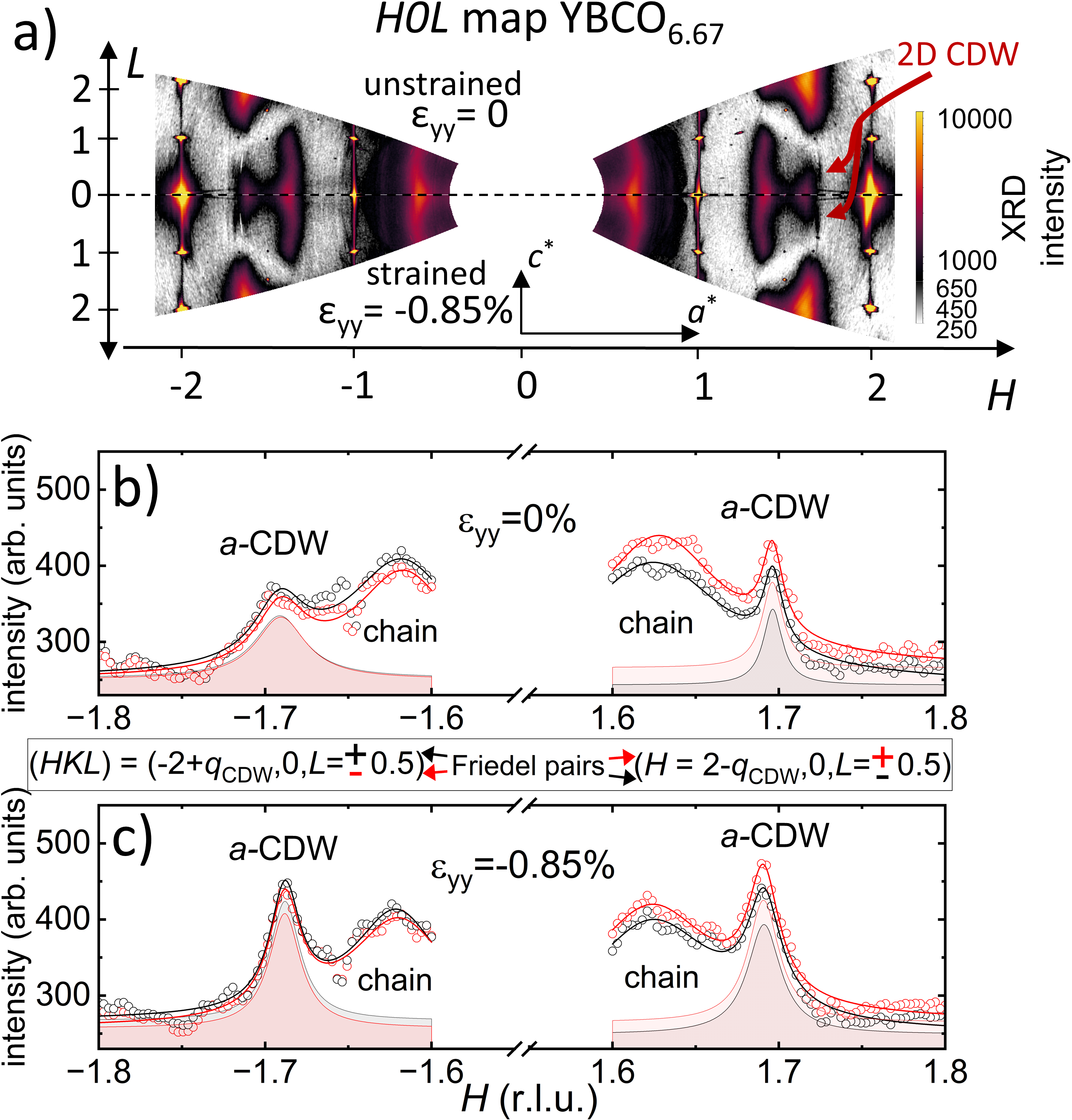}
  \caption{\label{2D_a_CDW_6p67} YBCO$_{6.67}$ ($H0L$) reciprocal space maps at zero (top) and highest $b$ axis compression (bottom) measured at $T_{c,0}$=67~K, show faint 2D $a$-CDW intensity at half-integer $L$ close to chain peaks that dominate the $H0L$ map. b) $H$-cuts across the 2D $a$-CDW of the unstrained map with anomalous widths for peaks at $H=-2+q_{\rm{CDW}}$ and their Friedel pairs at $2-q_{\rm{CDW}}$. In each case cuts for $L=\pm$0.5 are shown and Friedel pairs have the same color (red or black).c) Corresponding $H$ cuts through 2D $a$-CDW peaks at high $b$ axis compression, with similar peaks for both Friedel pairs.}
\end{figure}
In Supplementary Fig.\ref{2D_a_CDW_6p67}b we observe that the peak widths and intensities of the unstrained 2D $a$-CDW are anomalous. This can be seen from the asymmetric properties of Friedel pairs (peaks related by inversion symmetry). Comparing line cuts in panels b) and c) of Supplementary Fig.~\ref{2D_a_CDW_6p67} one can see that mirror symmetries are re-established and anomalous behavior of Friedel pairs is nearly suppressed at high strain, which potentially modifies the mosaicity of the sample. As the asymmetry is much reduced after compressing the sample, it is not due to an absorption effect but could be related to extinction.  Without providing any derivation, we observe a purely heuristic scaling between the square root of the extinction with the difference of the widths of 2D $a$-CDW Friedel pairs, $\sqrt{\textrm{extinction}}\propto \Delta w$ where $\Delta w$ is the difference of the widths of the a-CDW peaks, shown in Supplementary Fig.~\ref{2D_a_CDW_6p67}b,c. 
For small strains the small and large widths correspond to the correlations lengths $\xi_{a,>} =110 \pm 18 \angstrom$ and $\xi_{a,<} =47 \pm 7 \angstrom$. $\xi_{a,<}$ compares better to the longitudinal $a$-CDW correlation length of $41 \pm 1.7 \angstrom$ in ref.~\cite{Kim2021} and follows the expectations that $\xi_a$ should increase for $b$ axis compression. It is unclear why $a$-CDW peaks should have different widths, so the associated correlation lengths should be treated with caution.\\
Similar anomalous effects are also seen for the 2D $a$-CDW of the measured YBCO$_{6.80}$ sample. However, we focus on the 2D $b$-CDW, which is expected to increase with $a$ axis compression. In the $(0KL)$ map of Supplementary Fig.~\ref{O-III}b we observe no visible effect of the strain on the $b$-CDW. Consequently, the strain sensitivity must be reduced at this doping, although the negative pressure derivative, $\dfrac{\textrm{d}T_c}{\textrm{d}p_a}<0$, is suggestive of the contrary~\cite{Kraut1993}. Possibly, the higher disorder at this doping is problematic and the 3D CDW is reached only if the superconducting $T_c$ is reduced to values comparable to the onset temperature of the negative Hall number ($T_0 (p=0.14) \sim 50$~K) in high field experiments~\cite{LeBoeuf2011}.    
\begin{figure*} 
  \includegraphics[width=16cm]{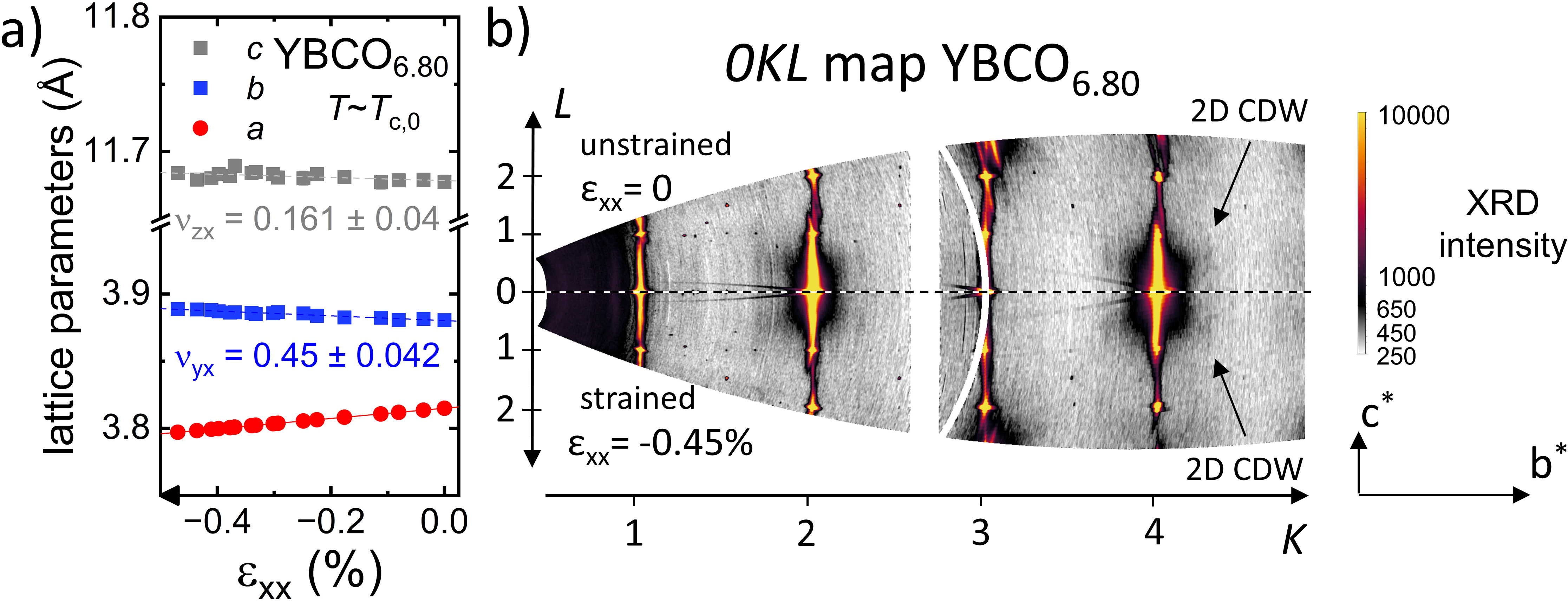}
  \caption{\label{O-III} (a) Strain evolution of the unit cell parameters for $a$ axis compression of the YBCO$_{6.80}$ sample. (b) YBCO$_{6.80}$ ($0KL$) reciprocal space maps at zero (top) and highest $a$ axis compression (bottom) measured at 77~K, just below $T_{c,0}$=78~K, show no visible effect from the compression at the wave vector of the 2D CDW. The lattice peaks are elongated due to worse sample quality, but this effect is strongly exaggerated due to the chosen upper cut-off when plotting the XRD intensity.}
\end{figure*}
\subsection{Structural information YBCO$_{\textbf{6.67}}$}
\begin{figure}
\hspace*{-0.3cm} 
  \includegraphics[width=8cm]{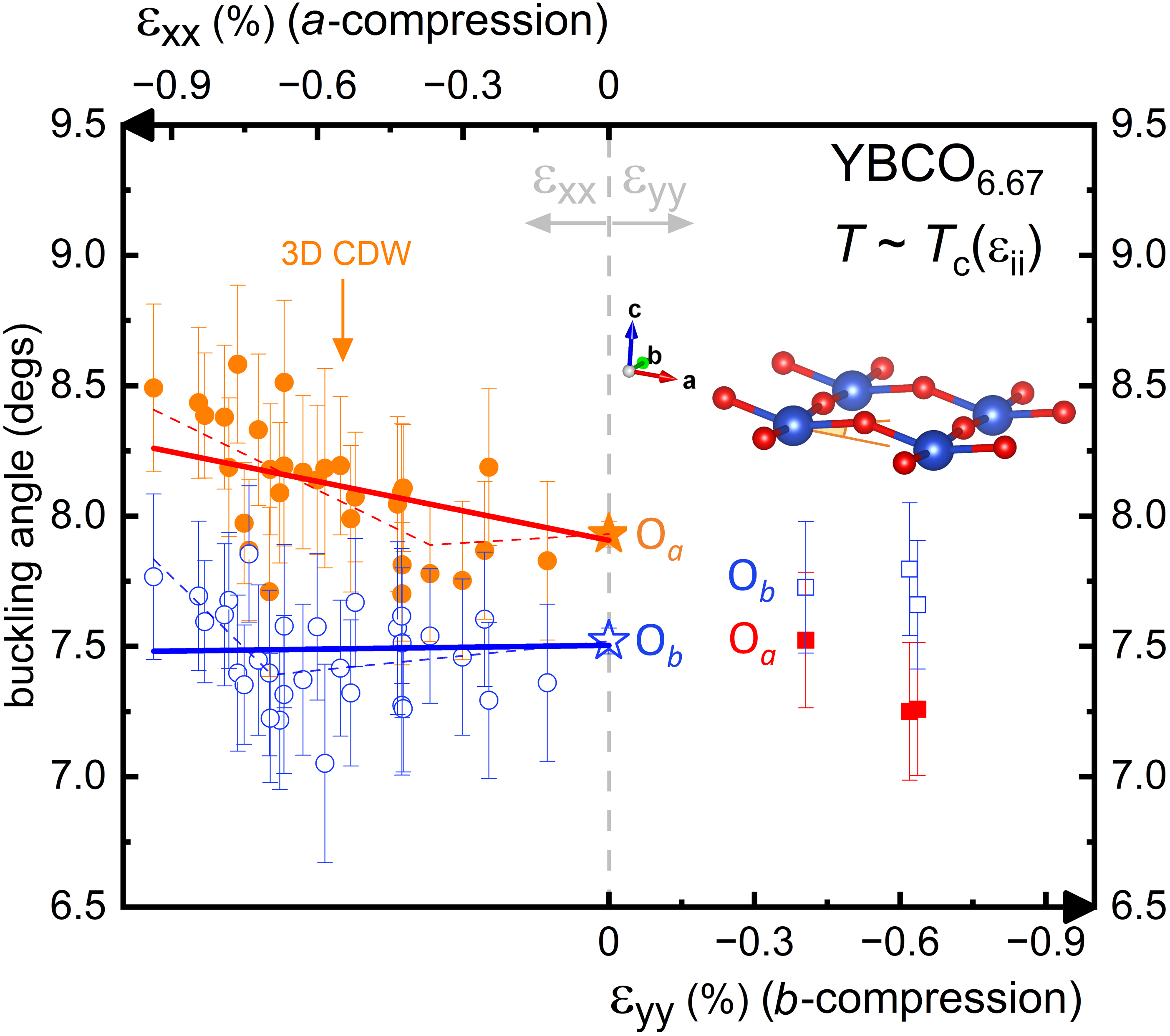}
  \caption{\label{buckling} (a) Planar oxygen bond buckling angles for $a$ and $b$ axis compression. $a$-compression amplifies the buckling of the bond along the $a$ axis (filled circles). The thick line is a linear fit. The dashed line is a piece-wise linear fit that has a kink at finite strain. For O$_a$ the kink is consistent within experimental uncertainty to the onset of 3D ordering (arrow). The bond along the $b$ axis (empty circles) is less sensitive to compression when the 3D CDW sets in. Stars refer to single crystal x-ray diffraction at 80~K (see text). The inset shows a CuO$_2$ plane with the buckling angle along the $a$ axis marked by the orange angle.}
\end{figure}
Refinements of the crystal structure with Jana2020 work best in the geometry for which the transmission axis lies in parallel to an in-plane axis of the unrotated sample (see Supplementary Table~\ref{tab:sample info}). All atomic coordinates in the plane are constrained by symmetry, and in this geometry high $L$-values are accessible which determine the $z$-coordinates. As noted in the previous section, refinements of the average crystal structure suffer from anomalously high secondary extinction at low strains, but the extinction drops off rapidly for increasing compression. We cross-checked the refinement results with x-ray diffraction at zero strain of a YBCO$_{6.67}$ sample with lab-based XRD on a rotating anode RIGAKU Synergy-DW Mo/Ag system. Using Mo $K_{\alpha}$ radiation and a resolution of 0.56 \AA we measured a $T$-dependent series. For all temperature steps around 20000 Bragg peaks were used, significantly reducing the experimental uncertainty. Given the very small temperature dependence of the refined parameters between 150~K and the lowest accessible temperature of 80~K, we present the corresponding crystallographic data taken at 80~K together with the strain-dependent refinements at $T\sim T_{c}(\varepsilon_{ii})$ in Supplementary Table~\ref{tab:refinements}.   
In Supplementary Fig.~\ref{buckling} and Supplementary Fig.~\ref{apical} we display results of structural refinements where we find the largest strain induced changes in the planar oxygen atoms. In Supplementary Fig.~\ref{buckling} 1\textdegree~of buckling corresponds to a vertical displacement of about 3 pm, so the scales in Supplementary Fig.~\ref{buckling} and Fig.~\ref{apical}a are comparable. All refinements (except the cross-check at zero strain) were carried out with isotropic temperature factors and both planar oxygens, O(2) (O$_a$) and O(3) (O$_b$), were assumed to possess equal temperature factors, as shown in Supplementary Fig.~\ref{apical}b.\\
Finally, an interesting structural parameter to be observed under uniaxial strain is the position of the apical oxygen above the Cu bilayer (see Supplementary Fig.~\ref{apical}a). It strongly impacts the intralayer hopping and has been identified as an essential control parameter of the superconducting $T_c$ in cuprates~\cite{Pavarini_PRL2001}. For instance, the motion of apical oxygen along the $c$ axis, together with a suppression of buckling under intense THz laser pumping have been identified as structural signatures of the high temperature transient superconducting phase of the cuprates~\cite{Kaiser2017,Mankowsky_2014}.
Under uniaxial compression, we do not note any change of the apical oxygen position with respect to the CuO$_2$ plane, nor of the distance between the CuO$_2$ layers within the bilayers. This indicates that a potential structural contribution to the reduction of $T_c$ here is mainly controlled by the changes in the CuO$_2$ planes.  

\begin{figure*}
\hspace*{-0.3cm}    
  \includegraphics[width=15cm]{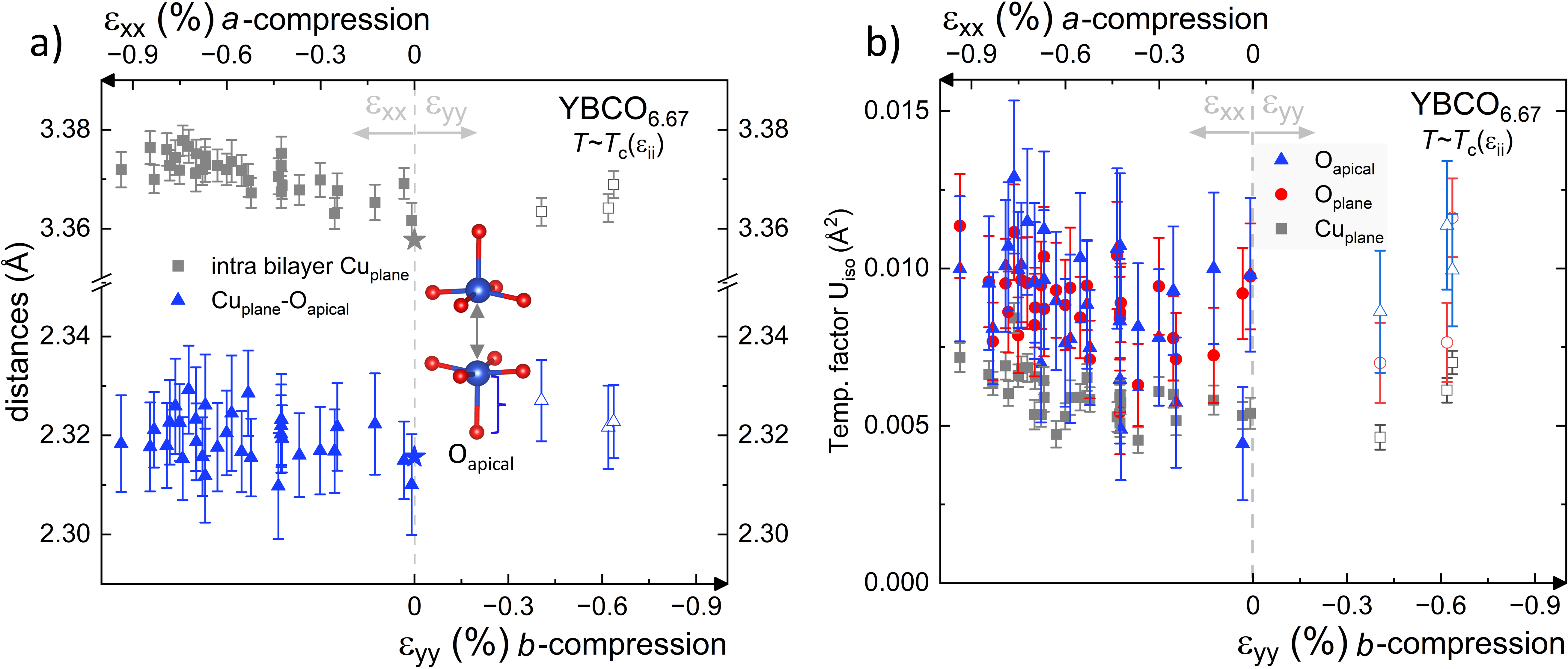}
  \caption{\label{apical} (a) Distances of the intra-bilayer distance of planar Cu atoms and the planar Cu to apical O situated above for $a$ and $b$ axis compression. While planar Cu is weakly affected, the apical oxygen does not move as 3D CDW ordering sets in around $\varepsilon_{xx}$=-0.55 \%. (b) Isotropic temperature factors of apical and planar oxygen and planar Cu, with no remarkable features larger than the statistical uncertainty.}
\end{figure*}
\begin{table}
\caption{\label{tab:refinements}%
 Refinement results for YBCO$_{6.67}$ at zero strain and 80~K, compared to $a$ axis compression $\varepsilon_{xx} = -0.845\%$ and 60~K. Due to fewer reflections, the strained samples were refined with isotropic temperature factors ($U_{\rm{iso}}$), whereas the unstrained reference measurement was refined with anisotropic temperature factors for which the equivalent isotropic temperature factor ($U_{\rm{eqiv}}$) is calculated.}
 \begin{ruledtabular}
\begin{tabular}{cccc}
 & YBa$_{2}$Cu$_{3}$O$_{\rm{6.67}}$ & unstrained (80 K) & strained (60 K) \\
\colrule
  & $a$ ($\angstrom$) & 3.8182(1) &  3.7883(2) \\
  & $b$ ($\angstrom$) & 3.8764(1) &  3.8914(28) \\    
  & $c$ ($\angstrom$) & 11.6706(2) &  11.6983(4) \\
Y  & $U_{\rm{eqiv}}$, $U_{\rm{iso}}$ ($\angstrom^{2}$) & 0.0032(1) &  0.0067(4) \\
Ba  & $z$ & 0.18701(1) &  0.18684(7) \\
  & $U_{\rm{eqiv}}$, $U_{\rm{iso}}$ ($\angstrom^{2}$) & 0.0042(1) &  0.0083(4) \\
Cu(1)  & $U_{\rm{eqiv}}$, $U_{\rm{iso}}$ ($\angstrom^{2}$) & 0.0042(1) &  0.0079(5) \\
Cu(2)  & $z$ & 0.35615(3) &  0.35569(14) \\
  & $U_{\rm{eqiv}}$, $U_{\rm{iso}}$ ($\angstrom^{2}$) & 0.0028(1) &  0.0066(4) \\
O(1)  & $U_{\rm{eqiv}}$, $U_{\rm{iso}}$ ($\angstrom^{2}$) & 0.0050(9) &  0.0222(58) \\
O(2)  & $z$ & 0.37893(17) &  0.37970(82) \\
  & $U_{\rm{eqiv}}$, $U_{\rm{iso}}$ ($\angstrom^{2}$) & 0.0049(3) &  0.0096(15) \\ 
O(3)  & $z$ & 0.37807(16) &  0.37816(84) \\
  & $U_{\rm{eqiv}}$, $U_{\rm{iso}}$ ($\angstrom^{2}$) & 0.0041(3) &  0.0096(15) \\
O(4)  & $z$ & 0.15773(15) &  0.15757(75) \\
  & $U_{\rm{eqiv}}$, $U_{\rm{iso}}$ ($\angstrom^{2}$) & 0.0072(4) &  0.0095(21) \\
  & GOF & 1.24 &  3.19 \\
    & $wR_2$ (\%) & 3.15 &  9.93 \\
    & $R_1$ (\%) & 0.98 &  4.08 \\
    & extinction & 0.025(1) &  0.150(30) \\
\end{tabular}
\end{ruledtabular}
\end{table}
\clearpage


\bibliography{strainbib}